\documentclass[sigplan,nonacm,screen]{acmart}

\AtBeginDocument{%
  }

\usepackage{isabelle}
\usepackage{isabellesym}
\usepackage{listings}
\usepackage{balance}

\newcommand{\snip}[4]{\expandafter\newcommand\csname #1\endcsname{#4}}
\snip{onelessdegree}{1}{2}{%
\isacommand{definition}\isamarkupfalse%
\ one{\isacharunderscore}{\kern0pt}less{\isacharunderscore}{\kern0pt}degree{\isacharcolon}{\kern0pt}{\isacharcolon}{\kern0pt}\ {\isachardoublequoteopen}rmpoly\ {\isasymRightarrow}\ rmpoly{\isachardoublequoteclose}\isanewline
\ \ \isakeyword{where}\ {\isachardoublequoteopen}one{\isacharunderscore}{\kern0pt}less{\isacharunderscore}{\kern0pt}degree\ p\ {\isacharequal}{\kern0pt}\isanewline
\ \ \ \ p\ {\isacharminus}{\kern0pt}\ monom\ {\isacharparenleft}{\kern0pt}lead{\isacharunderscore}{\kern0pt}coeff\ p{\isacharparenright}{\kern0pt}\ {\isacharparenleft}{\kern0pt}degree\ p{\isacharparenright}{\kern0pt}{\isachardoublequoteclose}%
}%endsnip
\snip{coeffonelessdegree}{1}{2}{%
\isacommand{lemma}\isamarkupfalse%
\ coeff{\isacharunderscore}{\kern0pt}one{\isacharunderscore}{\kern0pt}less{\isacharunderscore}{\kern0pt}degree{\isacharcolon}{\kern0pt}\isanewline
\ \isakeyword{assumes}\ {\isachardoublequoteopen}one{\isacharunderscore}{\kern0pt}less{\isacharunderscore}{\kern0pt}degree\ p\ {\isasymnoteq}\ {\isadigit{0}}{\isachardoublequoteclose}\isanewline
\ \isakeyword{assumes}\ {\isachardoublequoteopen}i\ {\isasymle}\ degree\ {\isacharparenleft}{\kern0pt}one{\isacharunderscore}{\kern0pt}less{\isacharunderscore}{\kern0pt}degree\ p{\isacharparenright}{\kern0pt}{\isachardoublequoteclose}\isanewline
\ \isakeyword{shows}\ {\isachardoublequoteopen}coeff\ p\ i\ {\isacharequal}{\kern0pt}\ coeff\ {\isacharparenleft}{\kern0pt}one{\isacharunderscore}{\kern0pt}less{\isacharunderscore}{\kern0pt}degree\ p{\isacharparenright}{\kern0pt}\ i{\isachardoublequoteclose}%
}%endsnip
\snip{polywithold}{1}{2}{%
\isacommand{lemma}\isamarkupfalse%
\ poly{\isacharunderscore}{\kern0pt}with{\isacharunderscore}{\kern0pt}one{\isacharunderscore}{\kern0pt}less{\isacharunderscore}{\kern0pt}degree{\isacharcolon}{\kern0pt}\isanewline
\ \ \isakeyword{assumes}\ {\isachardoublequoteopen}degree\ p\ {\isacharequal}{\kern0pt}\ d{\isachardoublequoteclose}\ {\isachardoublequoteopen}d\ {\isachargreater}{\kern0pt}\ {\isadigit{0}}{\isachardoublequoteclose}\isanewline
\ \ \isakeyword{shows}\ {\isachardoublequoteopen}poly\ p\ x\ {\isacharequal}{\kern0pt}\ \isanewline
\ \ {\isacharparenleft}{\kern0pt}{\isasymSum}i{\isasymle}degree\ {\isacharparenleft}{\kern0pt}one{\isacharunderscore}{\kern0pt}less{\isacharunderscore}{\kern0pt}degree\ p{\isacharparenright}{\kern0pt}{\isachardot}{\kern0pt}\ \isanewline
\ \ \ \ coeff\ {\isacharparenleft}{\kern0pt}one{\isacharunderscore}{\kern0pt}less{\isacharunderscore}{\kern0pt}degree\ p{\isacharparenright}{\kern0pt}\ i\ {\isacharasterisk}{\kern0pt}\ x\ {\isacharcircum}{\kern0pt}\ i{\isacharparenright}{\kern0pt}\ {\isacharplus}{\kern0pt}\ \isanewline
\ \ \ \ {\isacharparenleft}{\kern0pt}coeff\ p\ d{\isacharparenright}{\kern0pt}{\isacharasterisk}{\kern0pt}{\isacharparenleft}{\kern0pt}x{\isacharcircum}{\kern0pt}d{\isacharparenright}{\kern0pt}{\isachardoublequoteclose}%
}%endsnip
\snip{evalmpoly}{1}{2}{%
\isacommand{definition}\isamarkupfalse%
\ eval{\isacharunderscore}{\kern0pt}mpoly{\isacharcolon}{\kern0pt}{\isacharcolon}{\kern0pt}\ {\isachardoublequoteopen}real\ list\ {\isasymRightarrow}\ real\ mpoly\ {\isasymRightarrow}\ real{\isachardoublequoteclose}\isanewline
\isakeyword{where}\ {\isachardoublequoteopen}eval{\isacharunderscore}{\kern0pt}mpoly\ L\ p\ {\isacharequal}{\kern0pt}\ insertion\ {\isacharparenleft}{\kern0pt}nth{\isacharunderscore}{\kern0pt}default\ {\isadigit{0}}\ L{\isacharparenright}{\kern0pt}\ p{\isachardoublequoteclose}%
}%endsnip
\snip{evalmpolypoly}{1}{2}{%
\isacommand{definition}\isamarkupfalse%
\ eval{\isacharunderscore}{\kern0pt}mpoly{\isacharunderscore}{\kern0pt}poly{\isacharcolon}{\kern0pt}{\isacharcolon}{\kern0pt}\ {\isachardoublequoteopen}real\ list\ {\isasymRightarrow}\ rmpoly\ {\isasymRightarrow}\ real\ poly{\isachardoublequoteclose}\isanewline
\ \ \isakeyword{where}\ {\isachardoublequoteopen}eval{\isacharunderscore}{\kern0pt}mpoly{\isacharunderscore}{\kern0pt}poly\ L\ p\ {\isacharequal}{\kern0pt}\ map{\isacharunderscore}{\kern0pt}poly\ {\isacharparenleft}{\kern0pt}eval{\isacharunderscore}{\kern0pt}mpoly\ L{\isacharparenright}{\kern0pt}\ p{\isachardoublequoteclose}%
}%endsnip
\snip{mpolyeval}{1}{2}{%
\isacommand{lemma}\isamarkupfalse%
\ same{\isacharunderscore}{\kern0pt}evaluations{\isacharunderscore}{\kern0pt}same{\isacharunderscore}{\kern0pt}mpoly{\isacharcolon}{\kern0pt}\ \isanewline
\ \ \isakeyword{assumes}\ {\isachardoublequoteopen}{\isacharparenleft}{\kern0pt}{\isasymAnd}\ L{\isachardot}{\kern0pt}\ eval{\isacharunderscore}{\kern0pt}mpoly\ L\ p\ {\isacharequal}{\kern0pt}\ eval{\isacharunderscore}{\kern0pt}mpoly\ L\ q{\isacharparenright}{\kern0pt}{\isachardoublequoteclose}\isanewline
\ \ \isakeyword{shows}\ {\isachardoublequoteopen}p\ {\isacharequal}{\kern0pt}\ q{\isachardoublequoteclose}%
}%endsnip
\snip{reindexeduniveval}{1}{2}{%
\isacommand{lemma}\isamarkupfalse%
\ reindexed{\isacharunderscore}{\kern0pt}univ{\isacharunderscore}{\kern0pt}qs{\isacharunderscore}{\kern0pt}eval{\isacharcolon}{\kern0pt}\isanewline
\ \ \isakeyword{assumes}\ {\isachardoublequoteopen}univ{\isacharunderscore}{\kern0pt}qs\ {\isacharequal}{\kern0pt}\ univariate{\isacharunderscore}{\kern0pt}in\ qs\ {\isadigit{0}}{\isachardoublequoteclose}\ \isanewline
\ \ \isakeyword{assumes}\ {\isachardoublequoteopen}reindexed{\isacharunderscore}{\kern0pt}univ{\isacharunderscore}{\kern0pt}qs\ {\isacharequal}{\kern0pt}\ \isanewline
\ \ \ \ map\ {\isacharparenleft}{\kern0pt}map{\isacharunderscore}{\kern0pt}poly\ {\isacharparenleft}{\kern0pt}lowerPoly\ {\isadigit{0}}\ {\isadigit{1}}{\isacharparenright}{\kern0pt}{\isacharparenright}{\kern0pt}\ univ{\isacharunderscore}{\kern0pt}qs{\isachardoublequoteclose}\isanewline
\ \ \isakeyword{shows}\ {\isachardoublequoteopen}map\ {\isacharparenleft}{\kern0pt}eval{\isacharunderscore}{\kern0pt}mpoly\ {\isacharparenleft}{\kern0pt}x{\isacharhash}{\kern0pt}xs{\isacharparenright}{\kern0pt}{\isacharparenright}{\kern0pt}\ qs\ {\isacharequal}{\kern0pt}\isanewline
\ \ {\isacharparenleft}{\kern0pt}map\ {\isacharparenleft}{\kern0pt}{\isasymlambda}p{\isachardot}{\kern0pt}\ {\isacharparenleft}{\kern0pt}poly\ p\ x{\isacharparenright}{\kern0pt}{\isacharparenright}{\kern0pt}\isanewline
\ \ \ \ {\isacharparenleft}{\kern0pt}map\ {\isacharparenleft}{\kern0pt}{\isasymlambda}q{\isachardot}{\kern0pt}\ eval{\isacharunderscore}{\kern0pt}mpoly{\isacharunderscore}{\kern0pt}poly\ xs\ q{\isacharparenright}{\kern0pt}\ reindexed{\isacharunderscore}{\kern0pt}univ{\isacharunderscore}{\kern0pt}qs{\isacharparenright}{\kern0pt}{\isacharparenright}{\kern0pt}{\isachardoublequoteclose}%
}%endsnip
\snip{mpolytopolyalt}{1}{2}{%
\isacommand{definition}\isamarkupfalse%
\ mpoly{\isacharunderscore}{\kern0pt}to{\isacharunderscore}{\kern0pt}mpoly{\isacharunderscore}{\kern0pt}poly{\isacharunderscore}{\kern0pt}alt\ {\isacharcolon}{\kern0pt}{\isacharcolon}{\kern0pt}\ {\isachardoublequoteopen}nat\ {\isasymRightarrow}\ \isanewline
\ \ {\isacharprime}{\kern0pt}a\ {\isacharcolon}{\kern0pt}{\isacharcolon}{\kern0pt}\ comm{\isacharunderscore}{\kern0pt}ring{\isacharunderscore}{\kern0pt}{\isadigit{1}}\ mpoly\ {\isasymRightarrow}\ {\isacharprime}{\kern0pt}a\ mpoly\ poly{\isachardoublequoteclose}\isanewline
\ \ \isakeyword{where}\ {\isachardoublequoteopen}mpoly{\isacharunderscore}{\kern0pt}to{\isacharunderscore}{\kern0pt}mpoly{\isacharunderscore}{\kern0pt}poly{\isacharunderscore}{\kern0pt}alt\ x\ p\ {\isacharequal}{\kern0pt}\ \isanewline
\ \ \ {\isacharparenleft}{\kern0pt}{\isasymSum}i{\isasymin}{\isacharbraceleft}{\kern0pt}{\isadigit{0}}{\isachardot}{\kern0pt}{\isachardot}{\kern0pt}MPoly{\isacharunderscore}{\kern0pt}Type{\isachardot}{\kern0pt}degree\ p\ x{\isacharbraceright}{\kern0pt}\ {\isachardot}{\kern0pt}\isanewline
\ \ \ \ \ monom\ {\isacharparenleft}{\kern0pt}isolate{\isacharunderscore}{\kern0pt}variable{\isacharunderscore}{\kern0pt}sparse\ p\ x\ i{\isacharparenright}{\kern0pt}\ i{\isacharparenright}{\kern0pt}{\isachardoublequoteclose}%
}%endsnip
\snip{multivasuniv}{1}{2}{%
\isacommand{lemma}\isamarkupfalse%
\ multivar{\isacharunderscore}{\kern0pt}as{\isacharunderscore}{\kern0pt}univar{\isacharcolon}{\kern0pt}\ \isanewline
\ \ \isakeyword{shows}\ {\isachardoublequoteopen}mpoly{\isacharunderscore}{\kern0pt}to{\isacharunderscore}{\kern0pt}mpoly{\isacharunderscore}{\kern0pt}poly{\isacharunderscore}{\kern0pt}alt\ x\ p\ {\isacharequal}{\kern0pt}\ \isanewline
\ \ \ \ mpoly{\isacharunderscore}{\kern0pt}to{\isacharunderscore}{\kern0pt}mpoly{\isacharunderscore}{\kern0pt}poly\ x\ p{\isachardoublequoteclose}%
}%endsnip
\snip{lcassumpgeneration}{1}{2}{%
\isacommand{function}\isamarkupfalse%
\ lc{\isacharunderscore}{\kern0pt}assump{\isacharunderscore}{\kern0pt}generation{\isacharcolon}{\kern0pt}{\isacharcolon}{\kern0pt}\ {\isachardoublequoteopen}rmpoly\ {\isasymRightarrow}\ assumps\ {\isasymRightarrow}\ \isanewline
\ \ {\isacharparenleft}{\kern0pt}assumps\ {\isasymtimes}\ rmpoly{\isacharparenright}{\kern0pt}\ list{\isachardoublequoteclose}\isanewline
\ \ \isakeyword{where}\ {\isachardoublequoteopen}lc{\isacharunderscore}{\kern0pt}assump{\isacharunderscore}{\kern0pt}generation\ q\ assumps\ {\isacharequal}{\kern0pt}\ \isanewline
\ \ \ \ {\isacharparenleft}{\kern0pt}if\ q\ {\isacharequal}{\kern0pt}\ {\isadigit{0}}\ then\ {\isacharbrackleft}{\kern0pt}{\isacharparenleft}{\kern0pt}assumps{\isacharcomma}{\kern0pt}\ {\isadigit{0}}{\isacharparenright}{\kern0pt}{\isacharbrackright}{\kern0pt}\ else\isanewline
\ \ \ \ {\isacharparenleft}{\kern0pt}case\ {\isacharparenleft}{\kern0pt}lookup{\isacharunderscore}{\kern0pt}assump{\isacharunderscore}{\kern0pt}aux\ {\isacharparenleft}{\kern0pt}lead{\isacharunderscore}{\kern0pt}coeff\ q{\isacharparenright}{\kern0pt}\isanewline
\ \ \ \ \ \ \ assumps{\isacharparenright}{\kern0pt}\ of\isanewline
\ \ \ \ \ None\ {\isasymRightarrow}\ let\ \isanewline
\ \ \ \ \ \ \ zero\ {\isacharequal}{\kern0pt}\ lc{\isacharunderscore}{\kern0pt}assump{\isacharunderscore}{\kern0pt}generation\ {\isacharparenleft}{\kern0pt}one{\isacharunderscore}{\kern0pt}less{\isacharunderscore}{\kern0pt}degree\ q{\isacharparenright}{\kern0pt}\ \isanewline
\ \ \ \ \ \ \ \ \ \ {\isacharparenleft}{\kern0pt}{\isacharparenleft}{\kern0pt}lead{\isacharunderscore}{\kern0pt}coeff\ q{\isacharcomma}{\kern0pt}\ {\isacharparenleft}{\kern0pt}{\isadigit{0}}{\isacharcolon}{\kern0pt}{\isacharcolon}{\kern0pt}rat{\isacharparenright}{\kern0pt}{\isacharparenright}{\kern0pt}\ {\isacharhash}{\kern0pt}\ assumps{\isacharparenright}{\kern0pt}{\isacharsemicolon}{\kern0pt}\isanewline
\ \ \ \ \ \ \ one\ {\isacharequal}{\kern0pt}\ {\isacharparenleft}{\kern0pt}{\isacharparenleft}{\kern0pt}lead{\isacharunderscore}{\kern0pt}coeff\ q{\isacharcomma}{\kern0pt}\ {\isacharparenleft}{\kern0pt}{\isadigit{1}}{\isacharcolon}{\kern0pt}{\isacharcolon}{\kern0pt}rat{\isacharparenright}{\kern0pt}{\isacharparenright}{\kern0pt}\ {\isacharhash}{\kern0pt}\ assumps{\isacharcomma}{\kern0pt}\ q{\isacharparenright}{\kern0pt}{\isacharsemicolon}{\kern0pt}\isanewline
\ \ \ \ \ \ \ minus{\isacharunderscore}{\kern0pt}one\ {\isacharequal}{\kern0pt}\ {\isacharparenleft}{\kern0pt}{\isacharparenleft}{\kern0pt}lead{\isacharunderscore}{\kern0pt}coeff\ q{\isacharcomma}{\kern0pt}\ {\isacharparenleft}{\kern0pt}{\isacharminus}{\kern0pt}{\isadigit{1}}{\isacharcolon}{\kern0pt}{\isacharcolon}{\kern0pt}rat{\isacharparenright}{\kern0pt}{\isacharparenright}{\kern0pt}\ {\isacharhash}{\kern0pt}\ assumps{\isacharcomma}{\kern0pt}\ q{\isacharparenright}{\kern0pt}\ \isanewline
\ \ \ \ \ \ \ in\ one{\isacharhash}{\kern0pt}minus{\isacharunderscore}{\kern0pt}one{\isacharhash}{\kern0pt}zero\isanewline
\ \ \ \ \ {\isacharbar}{\kern0pt}\ {\isacharparenleft}{\kern0pt}Some\ i{\isacharparenright}{\kern0pt}\ {\isasymRightarrow}\isanewline
\ \ \ \ \ \ \ \ {\isacharparenleft}{\kern0pt}if\ i\ {\isacharequal}{\kern0pt}\ {\isadigit{0}}\ then\ lc{\isacharunderscore}{\kern0pt}assump{\isacharunderscore}{\kern0pt}generation\ \isanewline
\ \ \ \ \ \ \ \ \ \ {\isacharparenleft}{\kern0pt}one{\isacharunderscore}{\kern0pt}less{\isacharunderscore}{\kern0pt}degree\ q{\isacharparenright}{\kern0pt}\ assumps\isanewline
\ \ \ \ \ \ \ \ else\ {\isacharbrackleft}{\kern0pt}{\isacharparenleft}{\kern0pt}assumps{\isacharcomma}{\kern0pt}\ q{\isacharparenright}{\kern0pt}{\isacharbrackright}{\kern0pt}{\isacharparenright}{\kern0pt}{\isacharparenright}{\kern0pt}{\isacharparenright}{\kern0pt}{\isachardoublequoteclose}%
}%endsnip
\snip{lcassumpgenerationlist}{1}{2}{%
\isacommand{fun}\isamarkupfalse%
\ lc{\isacharunderscore}{\kern0pt}assump{\isacharunderscore}{\kern0pt}generation{\isacharunderscore}{\kern0pt}list{\isacharcolon}{\kern0pt}{\isacharcolon}{\kern0pt}\ {\isachardoublequoteopen}rmpoly\ list\ {\isasymRightarrow}\ \isanewline
\ \ assumps\ {\isasymRightarrow}\ {\isacharparenleft}{\kern0pt}assumps\ {\isasymtimes}\ rmpoly\ list{\isacharparenright}{\kern0pt}\ list{\isachardoublequoteclose}\isanewline
\ \ \isakeyword{where}\ {\isachardoublequoteopen}lc{\isacharunderscore}{\kern0pt}assump{\isacharunderscore}{\kern0pt}generation{\isacharunderscore}{\kern0pt}list\ {\isacharbrackleft}{\kern0pt}{\isacharbrackright}{\kern0pt}\ assumps\ {\isacharequal}{\kern0pt}\ \isanewline
\ \ \ \ {\isacharbrackleft}{\kern0pt}{\isacharparenleft}{\kern0pt}assumps{\isacharcomma}{\kern0pt}\ {\isacharbrackleft}{\kern0pt}{\isacharbrackright}{\kern0pt}{\isacharparenright}{\kern0pt}{\isacharbrackright}{\kern0pt}{\isachardoublequoteclose}\isanewline
\ \ {\isacharbar}{\kern0pt}\ {\isachardoublequoteopen}lc{\isacharunderscore}{\kern0pt}assump{\isacharunderscore}{\kern0pt}generation{\isacharunderscore}{\kern0pt}list\ {\isacharparenleft}{\kern0pt}q{\isacharhash}{\kern0pt}qs{\isacharparenright}{\kern0pt}\ assumps\ {\isacharequal}{\kern0pt}\ \isanewline
\ \ \ \ {\isacharparenleft}{\kern0pt}let\ rec\ {\isacharequal}{\kern0pt}\ lc{\isacharunderscore}{\kern0pt}assump{\isacharunderscore}{\kern0pt}generation\ q\ assumps\ in\isanewline
\ \ \ \ concat\ {\isacharparenleft}{\kern0pt}map\ {\isacharparenleft}{\kern0pt}{\isasymlambda}{\isacharparenleft}{\kern0pt}new{\isacharunderscore}{\kern0pt}assumps{\isacharcomma}{\kern0pt}\ r{\isacharparenright}{\kern0pt}{\isachardot}{\kern0pt}\ {\isacharparenleft}{\kern0pt}let\ list{\isacharunderscore}{\kern0pt}rec\ {\isacharequal}{\kern0pt}\ \isanewline
\ \ \ \ \ \ lc{\isacharunderscore}{\kern0pt}assump{\isacharunderscore}{\kern0pt}generation{\isacharunderscore}{\kern0pt}list\ qs\ new{\isacharunderscore}{\kern0pt}assumps\ in\isanewline
\ \ \ \ \ \ map\ {\isacharparenleft}{\kern0pt}{\isasymlambda}elem{\isachardot}{\kern0pt}\ {\isacharparenleft}{\kern0pt}fst\ elem{\isacharcomma}{\kern0pt}\ r{\isacharhash}{\kern0pt}{\isacharparenleft}{\kern0pt}snd\ elem{\isacharparenright}{\kern0pt}{\isacharparenright}{\kern0pt}{\isacharparenright}{\kern0pt}\ \isanewline
\ \ \ \ \ \ list{\isacharunderscore}{\kern0pt}rec{\isacharparenright}{\kern0pt}{\isacharparenright}{\kern0pt}\ rec{\isacharparenright}{\kern0pt}{\isacharparenright}{\kern0pt}{\isachardoublequoteclose}%
}%endsnip
\snip{signdetermination}{1}{2}{%
\isacommand{fun}\isamarkupfalse%
\ sign{\isacharunderscore}{\kern0pt}determination{\isacharcolon}{\kern0pt}{\isacharcolon}{\kern0pt}\ {\isachardoublequoteopen}rmpoly\ list\ {\isasymRightarrow}\ assumps\ {\isasymRightarrow}\ {\isacharparenleft}{\kern0pt}assumps\ {\isasymtimes}\ rat\ list\ list{\isacharparenright}{\kern0pt}\ list{\isachardoublequoteclose}\isanewline
\ \ \isakeyword{where}\ {\isachardoublequoteopen}sign{\isacharunderscore}{\kern0pt}determination\ qs\ assumps\ {\isacharequal}{\kern0pt}\ \isanewline
\ \ {\isacharparenleft}{\kern0pt}let\ branches\ {\isacharequal}{\kern0pt}\ \isanewline
\ \ \ \ \ lc{\isacharunderscore}{\kern0pt}assump{\isacharunderscore}{\kern0pt}generation{\isacharunderscore}{\kern0pt}list\ qs\ assumps\ in\isanewline
\ \ \ concat\ {\isacharparenleft}{\kern0pt}map\ {\isacharparenleft}{\kern0pt}{\isasymlambda}branch{\isachardot}{\kern0pt}\ let\ \isanewline
\ \ \ \ \ \ poly{\isacharunderscore}{\kern0pt}p{\isacharunderscore}{\kern0pt}branch\ {\isacharequal}{\kern0pt}\ poly{\isacharunderscore}{\kern0pt}p{\isacharunderscore}{\kern0pt}in{\isacharunderscore}{\kern0pt}branch\ branch{\isacharsemicolon}{\kern0pt}\isanewline
\ \ \ \ \ \ {\isacharparenleft}{\kern0pt}pos{\isacharunderscore}{\kern0pt}limit{\isacharunderscore}{\kern0pt}branch{\isacharcomma}{\kern0pt}\ neg{\isacharunderscore}{\kern0pt}limit{\isacharunderscore}{\kern0pt}branch{\isacharparenright}{\kern0pt}\ {\isacharequal}{\kern0pt}\ \isanewline
\ \ \ \ \ \ \ \ limit{\isacharunderscore}{\kern0pt}points{\isacharunderscore}{\kern0pt}on{\isacharunderscore}{\kern0pt}branch\ branch{\isacharsemicolon}{\kern0pt}\isanewline
\ \ \ \ \ \ mat{\isacharunderscore}{\kern0pt}eq{\isacharunderscore}{\kern0pt}signs{\isacharunderscore}{\kern0pt}on{\isacharunderscore}{\kern0pt}branch\ {\isacharequal}{\kern0pt}\ extract{\isacharunderscore}{\kern0pt}signs\isanewline
\ \ \ \ \ \ \ \ {\isacharparenleft}{\kern0pt}calculate{\isacharunderscore}{\kern0pt}data{\isacharunderscore}{\kern0pt}assumps{\isacharunderscore}{\kern0pt}M\ poly{\isacharunderscore}{\kern0pt}p{\isacharunderscore}{\kern0pt}branch\ \isanewline
\ \ \ \ \ \ \ \ {\isacharparenleft}{\kern0pt}snd\ branch{\isacharparenright}{\kern0pt}\ {\isacharparenleft}{\kern0pt}fst\ branch{\isacharparenright}{\kern0pt}{\isacharparenright}{\kern0pt}\ in\ \isanewline
\ \ \ \ map\ {\isacharparenleft}{\kern0pt}{\isasymlambda}{\isacharparenleft}{\kern0pt}a{\isacharcomma}{\kern0pt}\ signs{\isacharparenright}{\kern0pt}{\isachardot}{\kern0pt}\ \isanewline
\ \ \ \ \ \ {\isacharparenleft}{\kern0pt}a{\isacharcomma}{\kern0pt}\ pos{\isacharunderscore}{\kern0pt}limit{\isacharunderscore}{\kern0pt}branch{\isacharhash}{\kern0pt}neg{\isacharunderscore}{\kern0pt}limit{\isacharunderscore}{\kern0pt}branch{\isacharhash}{\kern0pt}signs{\isacharparenright}{\kern0pt}{\isacharparenright}{\kern0pt}\ \isanewline
\ \ \ \ \ \ \ mat{\isacharunderscore}{\kern0pt}eq{\isacharunderscore}{\kern0pt}signs{\isacharunderscore}{\kern0pt}on{\isacharunderscore}{\kern0pt}branch{\isacharparenright}{\kern0pt}\ branches{\isacharparenright}{\kern0pt}{\isacharparenright}{\kern0pt}{\isachardoublequoteclose}%
}%endsnip
\snip{elimforall}{1}{2}{%
\isacommand{fun}\isamarkupfalse%
\ elim{\isacharunderscore}{\kern0pt}forall{\isacharcolon}{\kern0pt}{\isacharcolon}{\kern0pt}\ {\isachardoublequoteopen}atom\ fm\ {\isasymRightarrow}\ atom\ fm{\isachardoublequoteclose}\isanewline
\ \ \isakeyword{where}\ {\isachardoublequoteopen}elim{\isacharunderscore}{\kern0pt}forall\ F\ {\isacharequal}{\kern0pt}\ {\isacharparenleft}{\kern0pt}let\ \isanewline
\ \ \ \ qs\ {\isacharequal}{\kern0pt}\ extract{\isacharunderscore}{\kern0pt}polys\ F{\isacharsemicolon}{\kern0pt}\isanewline
\ \ \ \ univ{\isacharunderscore}{\kern0pt}qs\ {\isacharequal}{\kern0pt}\ univariate{\isacharunderscore}{\kern0pt}in\ qs\ {\isadigit{0}}{\isacharsemicolon}{\kern0pt}\isanewline
\ \ \ \ reindexed{\isacharunderscore}{\kern0pt}univ{\isacharunderscore}{\kern0pt}qs\ {\isacharequal}{\kern0pt}\ map\ \isanewline
\ \ \ \ \ \ {\isacharparenleft}{\kern0pt}map{\isacharunderscore}{\kern0pt}poly\ {\isacharparenleft}{\kern0pt}lowerPoly\ {\isadigit{0}}\ {\isadigit{1}}{\isacharparenright}{\kern0pt}{\isacharparenright}{\kern0pt}\ univ{\isacharunderscore}{\kern0pt}qs{\isacharsemicolon}{\kern0pt}\isanewline
\ \ \ \ initial{\isacharunderscore}{\kern0pt}data\ {\isacharequal}{\kern0pt}\ sign{\isacharunderscore}{\kern0pt}determination\isanewline
\ \ \ \ \ \ reindexed{\isacharunderscore}{\kern0pt}univ{\isacharunderscore}{\kern0pt}qs\ {\isacharbrackleft}{\kern0pt}{\isacharbrackright}{\kern0pt}{\isacharsemicolon}{\kern0pt}\isanewline
\ \ \ \ filtered{\isacharunderscore}{\kern0pt}data\ {\isacharequal}{\kern0pt}\ filter\ {\isacharparenleft}{\kern0pt}{\isasymlambda}{\isacharparenleft}{\kern0pt}assumps{\isacharcomma}{\kern0pt}\ signs{\isacharunderscore}{\kern0pt}list{\isacharparenright}{\kern0pt}{\isachardot}{\kern0pt}\isanewline
\ \ \ \ list{\isacharunderscore}{\kern0pt}all\ {\isacharparenleft}{\kern0pt}{\isasymlambda}\ signs{\isachardot}{\kern0pt}\ \isanewline
\ \ \ \ \ \ lookup{\isacharunderscore}{\kern0pt}sem{\isacharunderscore}{\kern0pt}M\ F\ {\isacharparenleft}{\kern0pt}zip\ qs\ signs{\isacharparenright}{\kern0pt}\ {\isacharequal}{\kern0pt}\ {\isacharparenleft}{\kern0pt}Some\ True{\isacharparenright}{\kern0pt}{\isacharparenright}{\kern0pt}\ \isanewline
\ \ \ \ signs{\isacharunderscore}{\kern0pt}list\isanewline
\ \ \ \ {\isacharparenright}{\kern0pt}\ initial{\isacharunderscore}{\kern0pt}data\isanewline
\ in\ create{\isacharunderscore}{\kern0pt}disjunction\ filtered{\isacharunderscore}{\kern0pt}data{\isacharparenright}{\kern0pt}{\isachardoublequoteclose}%
}%endsnip
\snip{elimforallCorrectOne}{1}{2}{%
\isacommand{lemma}\isamarkupfalse%
\ eval{\isacharunderscore}{\kern0pt}elim{\isacharunderscore}{\kern0pt}forall{\isacharunderscore}{\kern0pt}correct{\isadigit{1}}{\isacharcolon}{\kern0pt}\isanewline
\ \ \isakeyword{fixes}\ F{\isacharcolon}{\kern0pt}{\isacharcolon}{\kern0pt}\ {\isachardoublequoteopen}atom\ fm{\isachardoublequoteclose}\isanewline
\ \ \isakeyword{assumes}\ {\isachardoublequoteopen}countQuantifiers\ F\ {\isacharequal}{\kern0pt}\ {\isadigit{0}}{\isachardoublequoteclose}\isanewline
\ \ \isakeyword{assumes}\ {\isachardoublequoteopen}eval\ {\isacharparenleft}{\kern0pt}elim{\isacharunderscore}{\kern0pt}forall\ F{\isacharparenright}{\kern0pt}\ xs{\isachardoublequoteclose}\isanewline
\ \ \isakeyword{shows}\ {\isachardoublequoteopen}{\isacharparenleft}{\kern0pt}{\isasymforall}x{\isachardot}{\kern0pt}{\isacharparenleft}{\kern0pt}eval\ F\ {\isacharparenleft}{\kern0pt}x{\isacharhash}{\kern0pt}xs{\isacharparenright}{\kern0pt}{\isacharparenright}{\kern0pt}{\isacharparenright}{\kern0pt}{\isachardoublequoteclose}%
}%endsnip
\snip{elimforallCorrectTwo}{1}{2}{%
\isacommand{lemma}\isamarkupfalse%
\ eval{\isacharunderscore}{\kern0pt}elim{\isacharunderscore}{\kern0pt}forall{\isacharunderscore}{\kern0pt}correct{\isadigit{2}}{\isacharcolon}{\kern0pt}\isanewline
\ \ \isakeyword{fixes}\ F{\isacharcolon}{\kern0pt}{\isacharcolon}{\kern0pt}\ {\isachardoublequoteopen}atom\ fm{\isachardoublequoteclose}\isanewline
\ \ \isakeyword{assumes}\ {\isachardoublequoteopen}countQuantifiers\ F\ {\isacharequal}{\kern0pt}\ {\isadigit{0}}{\isachardoublequoteclose}\isanewline
\ \ \isakeyword{assumes}\ {\isachardoublequoteopen}{\isacharparenleft}{\kern0pt}{\isasymforall}x{\isachardot}{\kern0pt}\ {\isacharparenleft}{\kern0pt}eval\ F\ {\isacharparenleft}{\kern0pt}x{\isacharhash}{\kern0pt}xs{\isacharparenright}{\kern0pt}{\isacharparenright}{\kern0pt}{\isacharparenright}{\kern0pt}{\isachardoublequoteclose}\isanewline
\ \ \isakeyword{shows}\ {\isachardoublequoteopen}eval\ {\isacharparenleft}{\kern0pt}elim{\isacharunderscore}{\kern0pt}forall\ F{\isacharparenright}{\kern0pt}\ xs{\isachardoublequoteclose}%
}%endsnip
\snip{calculatedata}{1}{2}{%
\isacommand{fun}\isamarkupfalse%
\ calculate{\isacharunderscore}{\kern0pt}data{\isacharunderscore}{\kern0pt}assumps{\isacharunderscore}{\kern0pt}M{\isacharcolon}{\kern0pt}{\isacharcolon}{\kern0pt}\ {\isachardoublequoteopen}rmpoly\ {\isasymRightarrow}\ rmpoly\ list\isanewline
\ \ \ \ \ {\isasymRightarrow}\ assumps\ {\isasymRightarrow}\ {\isacharparenleft}{\kern0pt}assumps\ {\isasymtimes}\ matrix{\isacharunderscore}{\kern0pt}equation{\isacharparenright}{\kern0pt}\ list{\isachardoublequoteclose}\isanewline
\ \ \isakeyword{where}\isanewline
\ \ {\isachardoublequoteopen}calculate{\isacharunderscore}{\kern0pt}data{\isacharunderscore}{\kern0pt}assumps{\isacharunderscore}{\kern0pt}M\ p\ qs\ init{\isacharunderscore}{\kern0pt}assumps\ {\isacharequal}{\kern0pt}\ \isanewline
\ \ \ {\isacharparenleft}{\kern0pt}let\ len\ {\isacharequal}{\kern0pt}\ length\ qs\ in\isanewline
\ \ \ \ if\ len\ {\isacharequal}{\kern0pt}\ {\isadigit{0}}\ then\ map\ {\isacharparenleft}{\kern0pt}{\isasymlambda}{\isacharparenleft}{\kern0pt}assumps{\isacharcomma}{\kern0pt}{\isacharparenleft}{\kern0pt}a{\isacharcomma}{\kern0pt}{\isacharparenleft}{\kern0pt}b{\isacharcomma}{\kern0pt}c{\isacharparenright}{\kern0pt}{\isacharparenright}{\kern0pt}{\isacharparenright}{\kern0pt}{\isachardot}{\kern0pt}\ \isanewline
\ \ \ \ \ \ {\isacharparenleft}{\kern0pt}assumps{\isacharcomma}{\kern0pt}\ {\isacharparenleft}{\kern0pt}a{\isacharcomma}{\kern0pt}b{\isacharcomma}{\kern0pt}map\ {\isacharparenleft}{\kern0pt}drop\ {\isadigit{1}}{\isacharparenright}{\kern0pt}\ c{\isacharparenright}{\kern0pt}{\isacharparenright}{\kern0pt}{\isacharparenright}{\kern0pt}\ \isanewline
\ \ \ \ \ \ \ \ \ {\isacharparenleft}{\kern0pt}reduce{\isacharunderscore}{\kern0pt}system{\isacharunderscore}{\kern0pt}M\ p\ {\isacharbrackleft}{\kern0pt}{\isadigit{1}}{\isacharbrackright}{\kern0pt}\ \isanewline
\ \ \ \ \ \ \ \ \ {\isacharparenleft}{\kern0pt}base{\isacharunderscore}{\kern0pt}case{\isacharunderscore}{\kern0pt}info{\isacharunderscore}{\kern0pt}M{\isacharunderscore}{\kern0pt}assumps\ init{\isacharunderscore}{\kern0pt}assumps{\isacharparenright}{\kern0pt}{\isacharparenright}{\kern0pt}\isanewline
\ \ \ \ else\ if\ len\ {\isasymle}\ {\isadigit{1}}\ then\ reduce{\isacharunderscore}{\kern0pt}system{\isacharunderscore}{\kern0pt}M\ p\ qs\ \isanewline
\ \ \ \ \ \ {\isacharparenleft}{\kern0pt}base{\isacharunderscore}{\kern0pt}case{\isacharunderscore}{\kern0pt}info{\isacharunderscore}{\kern0pt}M{\isacharunderscore}{\kern0pt}assumps\ init{\isacharunderscore}{\kern0pt}assumps{\isacharparenright}{\kern0pt}\isanewline
\ \ \ \ else\isanewline
\ \ \ \ {\isacharparenleft}{\kern0pt}let\ q{\isadigit{1}}\ {\isacharequal}{\kern0pt}\ take\ {\isacharparenleft}{\kern0pt}len\ div\ {\isadigit{2}}{\isacharparenright}{\kern0pt}\ qs{\isacharsemicolon}{\kern0pt}\ \isanewline
\ \ \ \ \ \ \ \ \ left\ {\isacharequal}{\kern0pt}\ calculate{\isacharunderscore}{\kern0pt}data{\isacharunderscore}{\kern0pt}assumps{\isacharunderscore}{\kern0pt}M\ \isanewline
\ \ \ \ \ \ \ \ \ \ \ \ p\ q{\isadigit{1}}\ init{\isacharunderscore}{\kern0pt}assumps{\isacharsemicolon}{\kern0pt}\isanewline
\ \ \ \ \ \ \ \ \ q{\isadigit{2}}\ {\isacharequal}{\kern0pt}\ drop\ {\isacharparenleft}{\kern0pt}len\ div\ {\isadigit{2}}{\isacharparenright}{\kern0pt}\ qs{\isacharsemicolon}{\kern0pt}\ \isanewline
\ \ \ \ \ \ \ \ \ right\ {\isacharequal}{\kern0pt}\ calculate{\isacharunderscore}{\kern0pt}data{\isacharunderscore}{\kern0pt}assumps{\isacharunderscore}{\kern0pt}M\ \isanewline
\ \ \ \ \ \ \ \ \ \ \ \ p\ q{\isadigit{2}}\ init{\isacharunderscore}{\kern0pt}assumps{\isacharsemicolon}{\kern0pt}\isanewline
\ \ \ \ \ \ \ \ \ comb\ {\isacharequal}{\kern0pt}\ combine{\isacharunderscore}{\kern0pt}systems{\isacharunderscore}{\kern0pt}M\ p\ q{\isadigit{1}}\ left\ q{\isadigit{2}}\ right\ in\isanewline
\ \ \ \ \ \ \ \ \ reduce{\isacharunderscore}{\kern0pt}system{\isacharunderscore}{\kern0pt}M\ p\ {\isacharparenleft}{\kern0pt}fst\ comb{\isacharparenright}{\kern0pt}\ {\isacharparenleft}{\kern0pt}snd\ comb{\isacharparenright}{\kern0pt}{\isacharparenright}{\kern0pt}{\isacharparenright}{\kern0pt}{\isachardoublequoteclose}%
}%endsnip
\snip{qewithvs}{1}{2}{%
\isacommand{theorem}\isamarkupfalse%
\ qe{\isacharunderscore}{\kern0pt}withVS{\isacharunderscore}{\kern0pt}correct{\isacharcolon}{\kern0pt}\isanewline
\ \ \isakeyword{fixes}\ F{\isacharcolon}{\kern0pt}{\isacharcolon}{\kern0pt}\ {\isachardoublequoteopen}atom\ fm{\isachardoublequoteclose}\isanewline
\ \ \isakeyword{shows}\ {\isachardoublequoteopen}eval\ F\ xs\ {\isacharequal}{\kern0pt}\ \isanewline
\ \ \ \ \ \ \ \ \ eval\ {\isacharparenleft}{\kern0pt}qe{\isacharunderscore}{\kern0pt}with{\isacharunderscore}{\kern0pt}VS\ F{\isacharparenright}{\kern0pt}\ xs{\isachardoublequoteclose}\isanewline
\isadelimproof
\ \ %
\endisadelimproof
\isatagproof
\isacommand{using}\isamarkupfalse%
\ qe{\isacharunderscore}{\kern0pt}correct\ VSLEG\isanewline
\ \ \isacommand{by}\isamarkupfalse%
\ {\isacharparenleft}{\kern0pt}simp\ add{\isacharcolon}{\kern0pt}\ qe{\isacharunderscore}{\kern0pt}with{\isacharunderscore}{\kern0pt}VS{\isacharunderscore}{\kern0pt}def{\isacharparenright}{\kern0pt}%
\endisatagproof
{\isafoldproof}%
\isadelimproof
\endisadelimproof
}%endsnip
\snip{spmodsmultiv}{1}{2}{%
\isacommand{function}\isamarkupfalse%
\ spmods{\isacharunderscore}{\kern0pt}multiv{\isacharcolon}{\kern0pt}{\isacharcolon}{\kern0pt}\isanewline
\ \ {\isachardoublequoteopen}real\ mpoly\ poly\ {\isasymRightarrow}\isanewline
\ \ \ real\ mpoly\ poly\ {\isasymRightarrow}\isanewline
\ \ \ {\isacharparenleft}{\kern0pt}real\ mpoly\ {\isasymtimes}\ rat{\isacharparenright}{\kern0pt}\ list\ {\isasymRightarrow}\ \isanewline
\ \ \ {\isacharparenleft}{\kern0pt}{\isacharparenleft}{\kern0pt}real\ mpoly\ {\isasymtimes}\ rat{\isacharparenright}{\kern0pt}\ list\ {\isasymtimes}\ {\isacharparenleft}{\kern0pt}real\ mpoly\ poly\ list{\isacharparenright}{\kern0pt}{\isacharparenright}{\kern0pt}\ list{\isachardoublequoteclose}\ \isanewline
\ \ \isakeyword{where}\ {\isachardoublequoteopen}spmods{\isacharunderscore}{\kern0pt}multiv\ p\ q\ assumps\ {\isacharequal}{\kern0pt}\ {\isacharparenleft}{\kern0pt}\isanewline
\ \ if\ p\ {\isacharequal}{\kern0pt}\ {\isadigit{0}}\ then\ {\isacharbrackleft}{\kern0pt}{\isacharparenleft}{\kern0pt}assumps{\isacharcomma}{\kern0pt}{\isacharbrackleft}{\kern0pt}{\isacharbrackright}{\kern0pt}{\isacharparenright}{\kern0pt}{\isacharbrackright}{\kern0pt}\ else\isanewline
\ \ {\isacharparenleft}{\kern0pt}case\ {\isacharparenleft}{\kern0pt}lookup{\isacharunderscore}{\kern0pt}assump{\isacharunderscore}{\kern0pt}aux\ {\isacharparenleft}{\kern0pt}lead{\isacharunderscore}{\kern0pt}coeff\ p{\isacharparenright}{\kern0pt}\ assumps{\isacharparenright}{\kern0pt}\ of\isanewline
\ \ \ \ \ \ None\ {\isasymRightarrow}\ let\ \isanewline
\ \ \ \ \ \ \ \ left\ {\isacharequal}{\kern0pt}\ spmods{\isacharunderscore}{\kern0pt}multiv\ {\isacharparenleft}{\kern0pt}one{\isacharunderscore}{\kern0pt}less{\isacharunderscore}{\kern0pt}degree\ p{\isacharparenright}{\kern0pt}\ q\ \isanewline
\ \ \ \ \ \ \ \ \ \ {\isacharparenleft}{\kern0pt}{\isacharparenleft}{\kern0pt}lead{\isacharunderscore}{\kern0pt}coeff\ p{\isacharcomma}{\kern0pt}\ {\isacharparenleft}{\kern0pt}{\isadigit{0}}{\isacharcolon}{\kern0pt}{\isacharcolon}{\kern0pt}rat{\isacharparenright}{\kern0pt}{\isacharparenright}{\kern0pt}\ {\isacharhash}{\kern0pt}\ assumps{\isacharparenright}{\kern0pt}{\isacharsemicolon}{\kern0pt}\isanewline
\ \ \ \ \ \ \ \ right{\isacharunderscore}{\kern0pt}one\ {\isacharequal}{\kern0pt}\ spmods{\isacharunderscore}{\kern0pt}multiv{\isacharunderscore}{\kern0pt}aux\ p\ q\ \isanewline
\ \ \ \ \ \ \ \ \ \ {\isacharparenleft}{\kern0pt}{\isacharparenleft}{\kern0pt}lead{\isacharunderscore}{\kern0pt}coeff\ p{\isacharcomma}{\kern0pt}\ {\isacharparenleft}{\kern0pt}{\isadigit{1}}{\isacharcolon}{\kern0pt}{\isacharcolon}{\kern0pt}rat{\isacharparenright}{\kern0pt}{\isacharparenright}{\kern0pt}\ {\isacharhash}{\kern0pt}\ assumps{\isacharparenright}{\kern0pt}{\isacharsemicolon}{\kern0pt}\isanewline
\ \ \ \ \ \ \ \ right{\isacharunderscore}{\kern0pt}minus{\isacharunderscore}{\kern0pt}one\ {\isacharequal}{\kern0pt}\ spmods{\isacharunderscore}{\kern0pt}multiv{\isacharunderscore}{\kern0pt}aux\ p\ q\ \isanewline
\ \ \ \ \ \ \ \ \ \ {\isacharparenleft}{\kern0pt}{\isacharparenleft}{\kern0pt}lead{\isacharunderscore}{\kern0pt}coeff\ p{\isacharcomma}{\kern0pt}\ {\isacharparenleft}{\kern0pt}{\isacharminus}{\kern0pt}{\isadigit{1}}{\isacharcolon}{\kern0pt}{\isacharcolon}{\kern0pt}rat{\isacharparenright}{\kern0pt}{\isacharparenright}{\kern0pt}\ {\isacharhash}{\kern0pt}\ assumps{\isacharparenright}{\kern0pt}\ in\isanewline
\ \ \ \ \ \ \ \ left\ {\isacharat}{\kern0pt}\ {\isacharparenleft}{\kern0pt}right{\isacharunderscore}{\kern0pt}one\ {\isacharat}{\kern0pt}\ right{\isacharunderscore}{\kern0pt}minus{\isacharunderscore}{\kern0pt}one{\isacharparenright}{\kern0pt}\isanewline
\ \ \ \ \ \ {\isacharbar}{\kern0pt}\ {\isacharparenleft}{\kern0pt}Some\ i{\isacharparenright}{\kern0pt}\ {\isasymRightarrow}\isanewline
\ \ \ \ \ \ \ \ {\isacharparenleft}{\kern0pt}if\ i\ {\isacharequal}{\kern0pt}\ {\isadigit{0}}\ then\ spmods{\isacharunderscore}{\kern0pt}multiv\ {\isacharparenleft}{\kern0pt}one{\isacharunderscore}{\kern0pt}less{\isacharunderscore}{\kern0pt}degree\ p{\isacharparenright}{\kern0pt}\ q\ assumps\isanewline
\ \ \ \ \ \ \ \ else\ spmods{\isacharunderscore}{\kern0pt}multiv{\isacharunderscore}{\kern0pt}aux\ p\ q\ assumps{\isacharparenright}{\kern0pt}{\isacharparenright}{\kern0pt}{\isacharparenright}{\kern0pt}{\isachardoublequoteclose}%
}%endsnip
\snip{calcdatacorrect}{1}{2}{%
\isacommand{lemma}\isamarkupfalse%
\ multivariate{\isacharunderscore}{\kern0pt}calculate{\isacharunderscore}{\kern0pt}data{\isacharunderscore}{\kern0pt}correct{\isacharcolon}{\kern0pt}\isanewline
\ \ \isakeyword{assumes}\ mat{\isacharunderscore}{\kern0pt}eq{\isacharcolon}{\kern0pt}\ {\isachardoublequoteopen}{\isacharparenleft}{\kern0pt}assumps{\isacharcomma}{\kern0pt}\ mat{\isacharunderscore}{\kern0pt}eq{\isacharparenright}{\kern0pt}\ {\isasymin}\ \isanewline
\ \ \ \ set\ {\isacharparenleft}{\kern0pt}calculate{\isacharunderscore}{\kern0pt}data{\isacharunderscore}{\kern0pt}assumps{\isacharunderscore}{\kern0pt}M\ p\ qs\ init{\isacharunderscore}{\kern0pt}assumps{\isacharparenright}{\kern0pt}{\isachardoublequoteclose}\isanewline
\ \ \isakeyword{assumes}\ {\isachardoublequoteopen}{\isasymAnd}p\ n{\isachardot}{\kern0pt}\ {\isacharparenleft}{\kern0pt}p{\isacharcomma}{\kern0pt}n{\isacharparenright}{\kern0pt}\ {\isasymin}\ set\ assumps\ {\isasymLongrightarrow}\isanewline
\ \ \ \ satisfies{\isacharunderscore}{\kern0pt}evaluation\ val\ p\ n{\isachardoublequoteclose}\isanewline
\ \ \isakeyword{assumes}\ {\isachardoublequoteopen}eval{\isacharunderscore}{\kern0pt}p\ {\isacharequal}{\kern0pt}\ eval{\isacharunderscore}{\kern0pt}mpoly{\isacharunderscore}{\kern0pt}poly\ val\ p{\isachardoublequoteclose}\isanewline
\ \ \isakeyword{assumes}\ {\isachardoublequoteopen}eval{\isacharunderscore}{\kern0pt}qs\ {\isacharequal}{\kern0pt}\ map\ {\isacharparenleft}{\kern0pt}eval{\isacharunderscore}{\kern0pt}mpoly{\isacharunderscore}{\kern0pt}poly\ val{\isacharparenright}{\kern0pt}\ qs{\isachardoublequoteclose}\isanewline
\ \ \isakeyword{assumes}\ p{\isacharunderscore}{\kern0pt}nonzero{\isacharcolon}{\kern0pt}\ {\isachardoublequoteopen}eval{\isacharunderscore}{\kern0pt}mpoly{\isacharunderscore}{\kern0pt}poly\ val\ p\ {\isasymnoteq}\ {\isadigit{0}}{\isachardoublequoteclose}\isanewline
\ \ \isakeyword{shows}\ {\isachardoublequoteopen}calculate{\isacharunderscore}{\kern0pt}data{\isacharunderscore}{\kern0pt}R\ eval{\isacharunderscore}{\kern0pt}p\ eval{\isacharunderscore}{\kern0pt}qs\ {\isacharequal}{\kern0pt}\ mat{\isacharunderscore}{\kern0pt}eq{\isachardoublequoteclose}%
}%endsnip
\snip{calcdatacomplete}{1}{2}{%
\isacommand{lemma}\isamarkupfalse%
\ multivariate{\isacharunderscore}{\kern0pt}calculate{\isacharunderscore}{\kern0pt}data{\isacharunderscore}{\kern0pt}complete{\isacharcolon}{\kern0pt}\isanewline
\ \ \isakeyword{assumes}\ {\isachardoublequoteopen}{\isasymAnd}p\ n{\isachardot}{\kern0pt}\ {\isacharparenleft}{\kern0pt}p{\isacharcomma}{\kern0pt}n{\isacharparenright}{\kern0pt}\ {\isasymin}\ set\ init{\isacharunderscore}{\kern0pt}assumps\ {\isasymLongrightarrow}\ \isanewline
\ \ \ \ satisfies{\isacharunderscore}{\kern0pt}evaluation\ val\ p\ n{\isachardoublequoteclose}\isanewline
\ \ \isakeyword{shows}\ \ {\isachardoublequoteopen}{\isasymexists}\ {\isacharparenleft}{\kern0pt}assumps{\isacharcomma}{\kern0pt}\ mat{\isacharunderscore}{\kern0pt}eq{\isacharparenright}{\kern0pt}\ {\isasymin}\ \isanewline
\ \ \ \ set\ {\isacharparenleft}{\kern0pt}calculate{\isacharunderscore}{\kern0pt}data{\isacharunderscore}{\kern0pt}assumps{\isacharunderscore}{\kern0pt}M\ p\ qs\ init{\isacharunderscore}{\kern0pt}assumps{\isacharparenright}{\kern0pt}{\isachardot}{\kern0pt}\isanewline
\ \ \ \ {\isacharparenleft}{\kern0pt}{\isasymforall}\ {\isacharparenleft}{\kern0pt}p{\isacharcomma}{\kern0pt}n{\isacharparenright}{\kern0pt}\ {\isasymin}\ set\ assumps{\isachardot}{\kern0pt}\isanewline
\ \ \ \ \ satisfies{\isacharunderscore}{\kern0pt}evaluation\ val\ p\ n{\isacharparenright}{\kern0pt}{\isachardoublequoteclose}%
}%endsnip
\snip{getallvaluations}{1}{2}{%
\isacommand{lemma}\isamarkupfalse%
\ get{\isacharunderscore}{\kern0pt}all{\isacharunderscore}{\kern0pt}valuations{\isacharcolon}{\kern0pt}\isanewline
\ \ \isakeyword{assumes}\ {\isachardoublequoteopen}{\isasymAnd}p\ n{\isachardot}{\kern0pt}\ {\isacharparenleft}{\kern0pt}p{\isacharcomma}{\kern0pt}n{\isacharparenright}{\kern0pt}\ {\isasymin}\ set\ init{\isacharunderscore}{\kern0pt}assumps\ \isanewline
\ \ \ \ {\isasymLongrightarrow}\ satisfies{\isacharunderscore}{\kern0pt}evaluation\ val\ p\ n{\isachardoublequoteclose}\isanewline
\ \ \isakeyword{obtains}\ assumps\ signs\ \isakeyword{where}\ {\isachardoublequoteopen}{\isacharparenleft}{\kern0pt}assumps{\isacharcomma}{\kern0pt}\ signs{\isacharparenright}{\kern0pt}\ \isanewline
\ \ \ \ {\isasymin}\ set\ {\isacharparenleft}{\kern0pt}sign{\isacharunderscore}{\kern0pt}determination{\isacharunderscore}{\kern0pt}inner\ qs\ init{\isacharunderscore}{\kern0pt}assumps{\isacharparenright}{\kern0pt}{\isachardoublequoteclose}\isanewline
\ \ \ \ {\isachardoublequoteopen}{\isasymAnd}p\ n{\isachardot}{\kern0pt}\ {\isacharparenleft}{\kern0pt}p{\isacharcomma}{\kern0pt}n{\isacharparenright}{\kern0pt}\ {\isasymin}\ set\ assumps\ {\isasymLongrightarrow}\ \isanewline
\ \ \ \ satisfies{\isacharunderscore}{\kern0pt}evaluation\ val\ p\ n{\isachardoublequoteclose}%
}%endsnip
\snip{pulloutpolys}{1}{2}{%
\isacommand{lemma}\isamarkupfalse%
\ extract{\isacharunderscore}{\kern0pt}polys{\isacharunderscore}{\kern0pt}semantics{\isacharcolon}{\kern0pt}\isanewline
\ \ \isakeyword{assumes}\ {\isachardoublequoteopen}qs\ {\isacharequal}{\kern0pt}\ extract{\isacharunderscore}{\kern0pt}polys\ F{\isachardoublequoteclose}\isanewline
\ \ \isakeyword{assumes}\ {\isachardoublequoteopen}signs\ {\isacharequal}{\kern0pt}\ map\ {\isacharparenleft}{\kern0pt}mpoly{\isacharunderscore}{\kern0pt}sign\ val{\isacharparenright}{\kern0pt}\ qs{\isachardoublequoteclose}\isanewline
\ \ \isakeyword{assumes}\ {\isachardoublequoteopen}countQuantifiers\ F\ {\isacharequal}{\kern0pt}\ {\isadigit{0}}{\isachardoublequoteclose}\isanewline
\ \ \isakeyword{shows}\ {\isachardoublequoteopen}Some\ {\isacharparenleft}{\kern0pt}eval\ F\ val{\isacharparenright}{\kern0pt}\ {\isacharequal}{\kern0pt}\ lookup{\isacharunderscore}{\kern0pt}sem{\isacharunderscore}{\kern0pt}M\ F\ {\isacharparenleft}{\kern0pt}zip\ qs\ signs{\isacharparenright}{\kern0pt}{\isachardoublequoteclose}%
}%endsnip
\snip{multivTQone}{1}{2}{%
\isacommand{lemma}\isamarkupfalse%
\ multiv{\isacharunderscore}{\kern0pt}tarski{\isacharunderscore}{\kern0pt}query{\isacharunderscore}{\kern0pt}correct{\isacharcolon}{\kern0pt}\isanewline
\ \ \isakeyword{assumes}\ inset{\isacharcolon}{\kern0pt}\ {\isachardoublequoteopen}{\isacharparenleft}{\kern0pt}assumps{\isacharcomma}{\kern0pt}\ tarski{\isacharunderscore}{\kern0pt}query{\isacharparenright}{\kern0pt}\ {\isasymin}\ \isanewline
\ \ \ \ set\ {\isacharparenleft}{\kern0pt}construct{\isacharunderscore}{\kern0pt}NofI{\isacharunderscore}{\kern0pt}M\ p\ acc\ I\ J{\isacharparenright}{\kern0pt}{\isachardoublequoteclose}\isanewline
\ \ \isakeyword{assumes}\ val{\isacharcolon}{\kern0pt}\ {\isachardoublequoteopen}{\isasymAnd}f\ n{\isachardot}{\kern0pt}\ {\isacharparenleft}{\kern0pt}f{\isacharcomma}{\kern0pt}n{\isacharparenright}{\kern0pt}\ {\isasymin}\ set\ assumps\ {\isasymLongrightarrow}\ \isanewline
\ \ \ \ satisfies{\isacharunderscore}{\kern0pt}evaluation\ val\ f\ n{\isachardoublequoteclose}\isanewline
\ \ \isakeyword{shows}\ {\isachardoublequoteopen}tarski{\isacharunderscore}{\kern0pt}query\ {\isacharequal}{\kern0pt}\ construct{\isacharunderscore}{\kern0pt}NofI{\isacharunderscore}{\kern0pt}R\isanewline
\ \ \ \ {\isacharparenleft}{\kern0pt}eval{\isacharunderscore}{\kern0pt}mpoly{\isacharunderscore}{\kern0pt}poly\ val\ p{\isacharparenright}{\kern0pt}\ \isanewline
\ \ \ \ {\isacharparenleft}{\kern0pt}eval{\isacharunderscore}{\kern0pt}mpoly{\isacharunderscore}{\kern0pt}poly{\isacharunderscore}{\kern0pt}list\ val\ I{\isacharparenright}{\kern0pt}\ \isanewline
\ \ \ \ {\isacharparenleft}{\kern0pt}eval{\isacharunderscore}{\kern0pt}mpoly{\isacharunderscore}{\kern0pt}poly{\isacharunderscore}{\kern0pt}list\ val\ J{\isacharparenright}{\kern0pt}{\isachardoublequoteclose}%
}%endsnip
\snip{multivTQtwo}{1}{2}{%
\isacommand{lemma}\isamarkupfalse%
\ multiv{\isacharunderscore}{\kern0pt}tarski{\isacharunderscore}{\kern0pt}queries{\isacharunderscore}{\kern0pt}complete{\isacharcolon}{\kern0pt}\isanewline
\ \ \isakeyword{assumes}\ {\isachardoublequoteopen}{\isasymAnd}f\ n{\isachardot}{\kern0pt}\ {\isacharparenleft}{\kern0pt}f{\isacharcomma}{\kern0pt}n{\isacharparenright}{\kern0pt}\ {\isasymin}\ set\ init{\isacharunderscore}{\kern0pt}assumps\ {\isasymLongrightarrow}\ \isanewline
\ \ \ \ satisfies{\isacharunderscore}{\kern0pt}evaluation\ val\ f\ n{\isachardoublequoteclose}\isanewline
\ \ \isakeyword{shows}\ \ {\isachardoublequoteopen}{\isasymexists}\ {\isacharparenleft}{\kern0pt}assumps{\isacharcomma}{\kern0pt}\ tq{\isacharparenright}{\kern0pt}\ {\isasymin}\ \isanewline
\ \ \ \ set\ {\isacharparenleft}{\kern0pt}construct{\isacharunderscore}{\kern0pt}NofI{\isacharunderscore}{\kern0pt}M\ p\ init{\isacharunderscore}{\kern0pt}assumps\ I\ J{\isacharparenright}{\kern0pt}{\isachardot}{\kern0pt}\isanewline
\ \ \ {\isacharparenleft}{\kern0pt}{\isasymforall}{\isacharparenleft}{\kern0pt}p{\isacharcomma}{\kern0pt}n{\isacharparenright}{\kern0pt}{\isasymin}set\ assumps{\isachardot}{\kern0pt}\ satisfies{\isacharunderscore}{\kern0pt}evaluation\ val\ p\ n{\isacharparenright}{\kern0pt}{\isachardoublequoteclose}%
}%endsnip
\snip{qe}{1}{2}{%
\isacommand{fun}\isamarkupfalse%
\ qe{\isacharcolon}{\kern0pt}{\isacharcolon}{\kern0pt}\ {\isachardoublequoteopen}atom\ fm\ {\isasymRightarrow}\ atom\ fm{\isachardoublequoteclose}\isanewline
\ \ \isakeyword{where}\ \isanewline
\ \ \ \ {\isachardoublequoteopen}qe\ TrueF\ {\isacharequal}{\kern0pt}\ TrueF{\isachardoublequoteclose}\isanewline
\ \ {\isacharbar}{\kern0pt}\ {\isachardoublequoteopen}qe\ FalseF\ {\isacharequal}{\kern0pt}\ FalseF{\isachardoublequoteclose}\isanewline
\ \ {\isacharbar}{\kern0pt}\ {\isachardoublequoteopen}qe\ {\isacharparenleft}{\kern0pt}Atom\ a{\isacharparenright}{\kern0pt}\ {\isacharequal}{\kern0pt}\ {\isacharparenleft}{\kern0pt}Atom\ a{\isacharparenright}{\kern0pt}{\isachardoublequoteclose}\isanewline
\ \ {\isacharbar}{\kern0pt}\ {\isachardoublequoteopen}qe\ {\isacharparenleft}{\kern0pt}And\ F{\isadigit{1}}\ F{\isadigit{2}}{\isacharparenright}{\kern0pt}\ {\isacharequal}{\kern0pt}\ And\ {\isacharparenleft}{\kern0pt}qe\ F{\isadigit{1}}{\isacharparenright}{\kern0pt}\ {\isacharparenleft}{\kern0pt}qe\ F{\isadigit{2}}{\isacharparenright}{\kern0pt}{\isachardoublequoteclose}\isanewline
\ \ {\isacharbar}{\kern0pt}\ {\isachardoublequoteopen}qe\ {\isacharparenleft}{\kern0pt}Or\ F{\isadigit{1}}\ F{\isadigit{2}}{\isacharparenright}{\kern0pt}\ {\isacharequal}{\kern0pt}\ Or\ {\isacharparenleft}{\kern0pt}qe\ F{\isadigit{1}}{\isacharparenright}{\kern0pt}\ {\isacharparenleft}{\kern0pt}qe\ F{\isadigit{2}}{\isacharparenright}{\kern0pt}{\isachardoublequoteclose}\isanewline
\ \ {\isacharbar}{\kern0pt}\ {\isachardoublequoteopen}qe\ {\isacharparenleft}{\kern0pt}Neg\ F{\isacharparenright}{\kern0pt}\ {\isacharequal}{\kern0pt}\ Neg\ {\isacharparenleft}{\kern0pt}qe\ F{\isacharparenright}{\kern0pt}{\isachardoublequoteclose}\isanewline
\ \ {\isacharbar}{\kern0pt}\ {\isachardoublequoteopen}qe\ {\isacharparenleft}{\kern0pt}ExQ\ F{\isacharparenright}{\kern0pt}\ {\isacharequal}{\kern0pt}\ elim{\isacharunderscore}{\kern0pt}exist\ {\isacharparenleft}{\kern0pt}qe\ F{\isacharparenright}{\kern0pt}{\isachardoublequoteclose}\isanewline
\ \ {\isacharbar}{\kern0pt}\ {\isachardoublequoteopen}qe\ {\isacharparenleft}{\kern0pt}AllQ\ F{\isacharparenright}{\kern0pt}\ {\isacharequal}{\kern0pt}\ elim{\isacharunderscore}{\kern0pt}forall\ {\isacharparenleft}{\kern0pt}qe\ F{\isacharparenright}{\kern0pt}{\isachardoublequoteclose}\isanewline
\ \ {\isacharbar}{\kern0pt}\ {\isachardoublequoteopen}qe\ {\isacharparenleft}{\kern0pt}AllN\ n\ F{\isacharparenright}{\kern0pt}\ {\isacharequal}{\kern0pt}\ {\isacharparenleft}{\kern0pt}elim{\isacharunderscore}{\kern0pt}forall\ {\isacharcircum}{\kern0pt}{\isacharcircum}{\kern0pt}\ n{\isacharparenright}{\kern0pt}\ {\isacharparenleft}{\kern0pt}qe\ F{\isacharparenright}{\kern0pt}{\isachardoublequoteclose}\isanewline
\ \ {\isacharbar}{\kern0pt}\ {\isachardoublequoteopen}qe\ {\isacharparenleft}{\kern0pt}ExN\ n\ F{\isacharparenright}{\kern0pt}\ {\isacharequal}{\kern0pt}\ {\isacharparenleft}{\kern0pt}elim{\isacharunderscore}{\kern0pt}exist\ {\isacharcircum}{\kern0pt}{\isacharcircum}{\kern0pt}\ n{\isacharparenright}{\kern0pt}\ {\isacharparenleft}{\kern0pt}qe\ F{\isacharparenright}{\kern0pt}{\isachardoublequoteclose}%
}%endsnip
\snip{qecorrect}{1}{2}{%
\isacommand{theorem}\isamarkupfalse%
\ qe{\isacharunderscore}{\kern0pt}correct{\isacharcolon}{\kern0pt}\isanewline
\ \isakeyword{fixes}\ F{\isacharcolon}{\kern0pt}{\isacharcolon}{\kern0pt}\ {\isachardoublequoteopen}atom\ fm{\isachardoublequoteclose}\isanewline
\ \isakeyword{shows}\ {\isachardoublequoteopen}eval\ F\ {\isasymnu}\ {\isacharequal}{\kern0pt}\ eval\ {\isacharparenleft}{\kern0pt}qe\ F{\isacharparenright}{\kern0pt}\ {\isasymnu}{\isachardoublequoteclose}%
}%endsnip
\snip{qeremoves}{1}{2}{%
\isacommand{theorem}\isamarkupfalse%
\ qe{\isacharunderscore}{\kern0pt}complete{\isacharcolon}{\kern0pt}\isanewline
\ \ \isakeyword{shows}\ {\isachardoublequoteopen}countQuantifiers\ {\isacharparenleft}{\kern0pt}qe\ F{\isacharparenright}{\kern0pt}\ {\isacharequal}{\kern0pt}\ {\isadigit{0}}{\isachardoublequoteclose}%
}%endsnip
\snip{spmodsauxdef}{1}{2}{%
\isacommand{function}\isamarkupfalse%
\ spmods{\isacharunderscore}{\kern0pt}multiv{\isacharunderscore}{\kern0pt}aux{\isacharcolon}{\kern0pt}{\isacharcolon}{\kern0pt}\isanewline
\ \ {\isachardoublequoteopen}rmpoly\ {\isasymRightarrow}\ rmpoly\ {\isasymRightarrow}\ assumps\ {\isasymRightarrow}\isanewline
\ \ {\isacharparenleft}{\kern0pt}assumps\ {\isasymtimes}\ rmpoly\ list{\isacharparenright}{\kern0pt}\ list{\isachardoublequoteclose}\ \isakeyword{where}\isanewline
\ \ {\isachardoublequoteopen}spmods{\isacharunderscore}{\kern0pt}multiv{\isacharunderscore}{\kern0pt}aux\ p\ q\ assumps\ {\isacharequal}{\kern0pt}\ {\isacharparenleft}{\kern0pt}\isanewline
\ \ if\ q\ {\isacharequal}{\kern0pt}\ {\isadigit{0}}\ then\ {\isacharbrackleft}{\kern0pt}{\isacharparenleft}{\kern0pt}assumps{\isacharcomma}{\kern0pt}\ {\isacharbrackleft}{\kern0pt}p{\isacharbrackright}{\kern0pt}{\isacharparenright}{\kern0pt}{\isacharbrackright}{\kern0pt}\isanewline
\ \ else\isanewline
\ \ case\ {\isacharparenleft}{\kern0pt}lookup{\isacharunderscore}{\kern0pt}assump{\isacharunderscore}{\kern0pt}aux\ {\isacharparenleft}{\kern0pt}lead{\isacharunderscore}{\kern0pt}coeff\ q{\isacharparenright}{\kern0pt}\ assumps{\isacharparenright}{\kern0pt}\ of\isanewline
\ \ None\ {\isasymRightarrow}\ let\ \isanewline
\ \ \ \ lcz\ {\isacharequal}{\kern0pt}\ spmods{\isacharunderscore}{\kern0pt}multiv{\isacharunderscore}{\kern0pt}aux\ p\ {\isacharparenleft}{\kern0pt}one{\isacharunderscore}{\kern0pt}less{\isacharunderscore}{\kern0pt}degree\ q{\isacharparenright}{\kern0pt}\isanewline
\ \ \ \ \ \ \ \ \ \ \ \ \ \ {\isacharparenleft}{\kern0pt}{\isacharparenleft}{\kern0pt}lead{\isacharunderscore}{\kern0pt}coeff\ q{\isacharcomma}{\kern0pt}\ {\isadigit{0}}{\isacharparenright}{\kern0pt}\ {\isacharhash}{\kern0pt}\ assumps{\isacharparenright}{\kern0pt}{\isacharsemicolon}{\kern0pt}\isanewline
\ \ \ \ lcp\ {\isacharequal}{\kern0pt}\ spmods{\isacharunderscore}{\kern0pt}multiv{\isacharunderscore}{\kern0pt}aux\ q\ {\isacharparenleft}{\kern0pt}mul{\isacharunderscore}{\kern0pt}pseudo{\isacharunderscore}{\kern0pt}mod\ p\ q{\isacharparenright}{\kern0pt}\isanewline
\ \ \ \ \ \ \ \ \ \ \ \ \ \ {\isacharparenleft}{\kern0pt}{\isacharparenleft}{\kern0pt}lead{\isacharunderscore}{\kern0pt}coeff\ q{\isacharcomma}{\kern0pt}\ {\isadigit{1}}{\isacharparenright}{\kern0pt}\ {\isacharhash}{\kern0pt}\ assumps{\isacharparenright}{\kern0pt}{\isacharsemicolon}{\kern0pt}\isanewline
\ \ \ \ lcn\ {\isacharequal}{\kern0pt}\ spmods{\isacharunderscore}{\kern0pt}multiv{\isacharunderscore}{\kern0pt}aux\ q\ {\isacharparenleft}{\kern0pt}mul{\isacharunderscore}{\kern0pt}pseudo{\isacharunderscore}{\kern0pt}mod\ p\ q{\isacharparenright}{\kern0pt}\isanewline
\ \ \ \ \ \ \ \ \ \ \ \ \ \ {\isacharparenleft}{\kern0pt}{\isacharparenleft}{\kern0pt}lead{\isacharunderscore}{\kern0pt}coeff\ q{\isacharcomma}{\kern0pt}\ {\isacharminus}{\kern0pt}{\isadigit{1}}{\isacharparenright}{\kern0pt}\ {\isacharhash}{\kern0pt}\ assumps{\isacharparenright}{\kern0pt}\ in\isanewline
\ \ \ \ \ \ {\isacharbrackleft}{\kern0pt}{\isacharbrackright}{\kern0pt}\isanewline
\ \ {\isacharbar}{\kern0pt}\ {\isacharparenleft}{\kern0pt}Some\ i{\isacharparenright}{\kern0pt}\ {\isasymRightarrow}\ {\isacharbrackleft}{\kern0pt}{\isacharbrackright}{\kern0pt}{\isacharparenright}{\kern0pt}{\isachardoublequoteclose}%
}%endsnip
\snip{pseudomod}{1}{2}{%
\isacommand{definition}\isamarkupfalse%
\ mul{\isacharunderscore}{\kern0pt}pseudo{\isacharunderscore}{\kern0pt}mod{\isacharcolon}{\kern0pt}{\isacharcolon}{\kern0pt}\ {\isachardoublequoteopen}{\isacharprime}{\kern0pt}a{\isacharcolon}{\kern0pt}{\isacharcolon}{\kern0pt}{\isacharbraceleft}{\kern0pt}comm{\isacharunderscore}{\kern0pt}ring{\isacharunderscore}{\kern0pt}{\isadigit{1}}{\isacharcomma}{\kern0pt}\isanewline
\ \ \ \ semiring{\isacharunderscore}{\kern0pt}{\isadigit{1}}{\isacharunderscore}{\kern0pt}no{\isacharunderscore}{\kern0pt}zero{\isacharunderscore}{\kern0pt}divisors{\isacharbraceright}{\kern0pt}\ Polynomial{\isachardot}{\kern0pt}poly\ {\isasymRightarrow}\ \isanewline
\ \ \ \ {\isacharprime}{\kern0pt}a\ Polynomial{\isachardot}{\kern0pt}poly\ {\isasymRightarrow}\ {\isacharprime}{\kern0pt}a\ Polynomial{\isachardot}{\kern0pt}poly{\isachardoublequoteclose}\ \isakeyword{where}\isanewline
\ \ {\isachardoublequoteopen}mul{\isacharunderscore}{\kern0pt}pseudo{\isacharunderscore}{\kern0pt}mod\ p\ q\ {\isacharequal}{\kern0pt}\ {\isacharparenleft}{\kern0pt}let\ m\ {\isacharequal}{\kern0pt}\isanewline
\ \ {\isacharparenleft}{\kern0pt}if\ even{\isacharparenleft}{\kern0pt}Polynomial{\isachardot}{\kern0pt}degree\ p{\isacharplus}{\kern0pt}{\isadigit{1}}{\isacharminus}{\kern0pt}Polynomial{\isachardot}{\kern0pt}degree\ q{\isacharparenright}{\kern0pt}\isanewline
\ \ \ \ \ \ then\ {\isacharminus}{\kern0pt}{\isadigit{1}}\ \ \ \ \ \ \ \ \isanewline
\ \ \ \ \ \ else\ {\isacharminus}{\kern0pt}Polynomial{\isachardot}{\kern0pt}lead{\isacharunderscore}{\kern0pt}coeff\ q{\isacharparenright}{\kern0pt}\ in\isanewline
\ \ \ \ Polynomial{\isachardot}{\kern0pt}smult\ m\ {\isacharparenleft}{\kern0pt}pseudo{\isacharunderscore}{\kern0pt}mod\ p\ q{\isacharparenright}{\kern0pt}{\isacharparenright}{\kern0pt}{\isachardoublequoteclose}%
}%endsnip

\usepackage{xspace}
\newcommand{\Isabelle}{Isabelle/HOL\xspace}
\newcommand{\bebecomes}{\mathrel{::=}}
\newcommand{\alternative}{~|~}

\usepackage{prettyref}
\newcommand{\rref}[2][]{\prettyref{#2}}
\newrefformat{sec}{Sec.\,\ref{#1}}
\newrefformat{subsec}{Sec.\,\ref{#1}}
\newrefformat{def}{Def.\,\ref{#1}}
\newrefformat{thm}{Theorem\,\ref{#1}}
\newrefformat{prop}{Proposition\,\ref{#1}}
\newrefformat{rem}{Remark\,\ref{#1}}
\newrefformat{lem}{Lemma\,\ref{#1}}
\newrefformat{cor}{Corollary\,\ref{#1}}
\newrefformat{ex}{Example\,\ref{#1}}
\newrefformat{cex}{Counterexample\,\ref{#1}}
\newrefformat{tab}{Table\,\ref{#1}}
\newrefformat{fig}{Fig.\,\ref{#1}}
\newrefformat{case}{case\,\ref{#1}}
\newrefformat{foot}{Footnote\,\ref{#1}}

\snip{spmodsauxmanual}{1}{2}{%
\isacommand{function}\isamarkupfalse%
\ spmods{\isacharunderscore}{\kern0pt}multiv{\isacharunderscore}{\kern0pt}aux{\isacharcolon}{\kern0pt}{\isacharcolon}{\kern0pt}\isanewline
\ \ {\isachardoublequoteopen}rmpoly\ {\isasymRightarrow}\ rmpoly\ {\isasymRightarrow}\ assumps\ {\isasymRightarrow}\isanewline
\ \ {\isacharparenleft}{\kern0pt}assumps\ {\isasymtimes}\ rmpoly\ list{\isacharparenright}{\kern0pt}\ list{\isachardoublequoteclose}\ \isakeyword{where}\isanewline
\ \ {\isachardoublequoteopen}spmods{\isacharunderscore}{\kern0pt}multiv{\isacharunderscore}{\kern0pt}aux\ p\ q\ assumps\ {\isacharequal}{\kern0pt}\ {\isacharparenleft}{\kern0pt}\isanewline
\ \ if\ q\ {\isacharequal}{\kern0pt}\ {\isadigit{0}}\ then\ {\isacharbrackleft}{\kern0pt}{\isacharparenleft}{\kern0pt}assumps{\isacharcomma}{\kern0pt}\ {\isacharbrackleft}{\kern0pt}p{\isacharbrackright}{\kern0pt}{\isacharparenright}{\kern0pt}{\isacharbrackright}{\kern0pt}\isanewline
\ \ else\isanewline
\ \ case\ {\isacharparenleft}{\kern0pt}lookup{\isacharunderscore}{\kern0pt}assump{\isacharunderscore}{\kern0pt}aux\ {\isacharparenleft}{\kern0pt}lead{\isacharunderscore}{\kern0pt}coeff\ q{\isacharparenright}{\kern0pt}\ assumps{\isacharparenright}{\kern0pt}\ of\isanewline
\ \ None\ {\isasymRightarrow}\isanewline
\ \ \ \ let\ lcz\ {\isacharequal}{\kern0pt}\ spmods{\isacharunderscore}{\kern0pt}multiv{\isacharunderscore}{\kern0pt}aux\ p\ {\isacharparenleft}{\kern0pt}one{\isacharunderscore}{\kern0pt}less{\isacharunderscore}{\kern0pt}degree\ q{\isacharparenright}{\kern0pt}\isanewline
\ \ \ \ \ \ \ \ \ \ \ \ \ \ {\isacharparenleft}{\kern0pt}{\isacharparenleft}{\kern0pt}lead{\isacharunderscore}{\kern0pt}coeff\ q{\isacharcomma}{\kern0pt}\ {\isadigit{0}}{\isacharparenright}{\kern0pt}\ {\isacharhash}{\kern0pt}\ assumps{\isacharparenright}{\kern0pt}\ in\isanewline
\ \ \ \ let\ lcp\ {\isacharequal}{\kern0pt}\ spmods{\isacharunderscore}{\kern0pt}multiv{\isacharunderscore}{\kern0pt}aux\ q\ {\isacharparenleft}{\kern0pt}mul{\isacharunderscore}{\kern0pt}pseudo{\isacharunderscore}{\kern0pt}mod\ p\ q{\isacharparenright}{\kern0pt}\isanewline
\ \ \ \ \ \ \ \ \ \ \ \ \ \ {\isacharparenleft}{\kern0pt}{\isacharparenleft}{\kern0pt}lead{\isacharunderscore}{\kern0pt}coeff\ q{\isacharcomma}{\kern0pt}\ {\isadigit{1}}{\isacharparenright}{\kern0pt}\ {\isacharhash}{\kern0pt}\ assumps{\isacharparenright}{\kern0pt}\ in\isanewline
\ \ \ \ let\ lcn\ {\isacharequal}{\kern0pt}\ spmods{\isacharunderscore}{\kern0pt}multiv{\isacharunderscore}{\kern0pt}aux\ q\ {\isacharparenleft}{\kern0pt}mul{\isacharunderscore}{\kern0pt}pseudo{\isacharunderscore}{\kern0pt}mod\ p\ q{\isacharparenright}{\kern0pt}\isanewline
\ \ \ \ \ \ \ \ \ \ \ \ \ \ {\isacharparenleft}{\kern0pt}{\isacharparenleft}{\kern0pt}lead{\isacharunderscore}{\kern0pt}coeff\ q{\isacharcomma}{\kern0pt}\ {\isacharminus}{\kern0pt}{\isadigit{1}}{\isacharparenright}{\kern0pt}\ {\isacharhash}{\kern0pt}\ assumps{\isacharparenright}{\kern0pt}\ in\isanewline
\ \ \ \ \ \ {\kern0pt}$\dots$  /* combine lcz, lcp, lcn */ {\kern0pt}\isanewline
\ \ {\isacharbar}{\kern0pt}\ {\isacharparenleft}{\kern0pt}Some\ i{\isacharparenright}{\kern0pt}\ {\isasymRightarrow}\ {\kern0pt} $\dots$ /* two recursive branches */ {\kern0pt}{\isacharparenright}{\kern0pt}{\isachardoublequoteclose}%
}%endsnip

\newtheorem*{remark}{Remark}

\begin{document}

\title[A First Complete Algorithm for Real Quantifier Elimination in Isabelle/HOL]{A First Complete Algorithm for\\ Real Quantifier Elimination in Isabelle/HOL}

\author{Katherine Kosaian}
\orcid{0000-0002-9336-6006}
\affiliation{%
   \institution{Carnegie Mellon University}
  \city{Pittsburgh}
   \state{PA}
  \country{USA}}
  \email{kcordwel@cs.cmu.edu}

\author{Yong Kiam Tan}
\orcid{0000-0001-7033-2463}
\affiliation{%
  \institution{Carnegie Mellon University}
  \city{Pittsburgh}
   \state{PA}
  \country{USA}}
  \email{yongkiat@alumni.cmu.edu}

\author{Andr\'{e} Platzer}
\orcid{0000-0001-7238-5710}
\affiliation{%
  \institution{Karlsruhe Institute of Technology}
  \city{Karlsruhe}
  \country{Germany}}
  \email{platzer@kit.edu}

\renewcommand{\shortauthors}{Katherine Kosaian, Yong Kiam Tan, and Andr\'{e} Platzer.}

\begin{abstract}
  We formalize a multivariate quantifier elimination (QE) algorithm in the theorem prover Isabelle/HOL.
 Our algorithm is complete, in that it is able to reduce \textit{any} quantified formula in the first-order logic of real arithmetic to a logically equivalent quantifier-free formula.
  The algorithm we formalize is a hybrid mixture of Tarski's original QE algorithm and the Ben-Or, Kozen, and Reif algorithm, and it is the first complete multivariate QE algorithm formalized in Isabelle/HOL.
  
\end{abstract}




\maketitle

\section{Introduction}
Real arithmetic problems appear in many application domains, including safety-critical application domains, such as the verification of cyber-physical systems (CPS).
Very often, these problems involve $\exists$ and $\forall$ quantifiers, which pose theoretical and practical computational challenges \cite{DBLP:journals/jsc/DavenportH88,DBLP:journals/jsc/Weispfenning88,DBLP:conf/cade/PlatzerQR09}.
The best known way of handling arbitrary quantified statements is with \textit{quantifier elimination (QE)}, which transforms quantified statements into logically equivalent quantifier-free formulas, which are then evaluated.
Alfred Tarski \cite{Tarski} proved that the theory of real-closed fields is decidable, by establishing that algorithms to perform quantifier elimination on formulas in the first-order logic of real arithmetic exist; in practice, these algorithms tend to be complicated.

Given the safety-critical nature of real arithmetic questions \cite{Platzer18}, it is not surprising that considerable attention has been given to formally verifying algorithms for real QE \cite{AssiaQE, BKR, DBLP:conf/cpp/Eberl15, DBLP:conf/tphol/Harrison07, DBLP:journals/mscs/Mahboubi07, li2019deciding, harrison, NASAHutch, NASATarski, DBLP:journals/jar/Nipkow10, DBLP:conf/cade/PlatzerQR09, scharager2021verified}.
However, while considerable progress has been made on verifying \textit{univariate} QE methods (methods for QE problems that only involve one variable, and so have at most one quantifier) \cite{BKR,DBLP:conf/cpp/Eberl15,li2019deciding,NASAHutch,NASATarski}, and while a variety of works have focused on verifying \textit{special-purpose} QE methods (that is, methods which target some fragment of multivariate QE problems) \cite{DBLP:conf/tphol/Harrison07,DBLP:journals/jar/Nipkow10,DBLP:conf/cade/PlatzerQR09,scharager2021verified}, only limited progress has been made on verifying \textit{complete} multivariate QE algorithms (i.e., algorithms that are capable of
resolving \emph{any} real QE problem).
Multivariate QE algorithms are significantly more challenging.
Multivariate polynomials are unlike univariate polynomials, because they may have infinitely many roots, their leading coefficients are themselves polynomials and may have zeros, polynomial division is not always unique, and ideal computations use Gr\"obner bases instead of Euclidian division.
Additionally, whereas univariate QE problems only involve a single quantifier and always reduce to True or False, multivariate QE problems can involve nested (alternating) quantifiers and free variables.

To our knowledge, the main published progress on verifying complete multivariate QE algorithms in theorem provers is threefold: first, Mahboubi \cite{DBLP:journals/mscs/Mahboubi07} \textit{implemented} (but did not yet verify) the fastest-known QE algorithm, \textit{cylindrical algebraic decomposition (CAD)} \cite{Collins} in Coq; second, McLaughlin and Harrison developed a \textit{proof-producing} (but not verified) procedure based on the Cohen-H\"{o}rmander algorithm in HOL Light \cite{harrison}; finally, Cohen and Mahboubi verified Tarski's original QE algorithm in Coq \cite{cohen_phd,AssiaQE}.
Unfortunately, both Tarski's original QE algorithm and the Cohen-H\"{o}rmander algorithm have non-elementary complexity (i.e. the complexity is not bounded by any tower of powers of two).
While McLaughlin and Harrison's procedure can solve simple microbenchmarks, they acknowledge considerable experimental limitations \cite{harrison}.\footnote{This is not only due to the complexity of the Cohen-H\"{o}rmander algorithm, but also because proof-producing algorithms are not verified once and for all but, instead, have to produce a new proof of correctness per question, which incurs significant overhead compared to fully verified ones \cite{harrison,DBLP:conf/cade/PlatzerQR09}.}
Similarly, Cohen and Mahboubi consider their work to be primarily a theoretical contribution \cite{AssiaQE}.

The dearth of efficient formally-verified support for QE is in part a consequence of the intricacy of QE algorithms.
There is arguably a tradeoff \cite{scharager2021verified} between the computational efficiency of an algorithm and the tractability of verification.
Most notably, the CAD algorithm is efficient but complex and tremendously difficult to verify; only the significantly simpler univariate case has been fully verified (independently, in Isabelle/HOL \cite{li2019deciding} and PVS \cite{NASAHutch}).
Further, in order for CAD to realize its full potential for efficiency, many further insights \cite{Brown,DBLP:journals/jsc/CollinsH91,DBLP:conf/issac/DolzmannSS04,McCallumProj} beyond the original development \cite{Collins} are needed, and improving CAD and algorithms for real QE at large is an active area of research.

The lack of efficient \emph{verified} QE methods is also a consequence of the challenge posed by verification.
Working within the formal setting of a theorem prover adds both a considerable layer of rigor but also intricacy, which is why even small progress needs significant effort.
For example, Mahboubi \cite{DBLP:journals/mscs/Mahboubi07} discusses the many challenges involved in implementing CAD in Coq---a significantly more arduous and involved task than implementing CAD in an unverified computer algebra system (which also took decades \cite{DBLP:journals/cca/Brown03,strzeMathematica}).

In our work, we target a potential \textit{sweet spot} within the tradeoff between complexity and verification amenability \cite{scharager2021verified} by verifying a \emph{complete} multivariate QE algorithm loosely based on the \textit{Ben-Or, Kozen, and Reif (BKR)} algorithm \cite{DBLP:journals/jcss/Ben-OrKR86} (but presently with less efficiency).
The BKR algorithm shares some theoretical similarity to Tarski's original QE algorithm (in that it uses a matrix equation to store sign information for polynomials), but it includes an additional reduction step for greater efficiency.
Although the multivariate complexity analysis in the paper describing BKR was flawed \cite{1993Improved}, rendering its stated bounds inaccurate, this was nevertheless an influential algorithm which was later extended into a number of improved and/or generalized variants with highly compelling parallel complexity bounds, including ones by Renegar \cite{DBLP:journals/jsc/Renegar92b}, Canny \cite{1993Improved}, and Cucker \emph{et al.} \cite{DBLP:journals/aaecc/CuckerLMPR92}.
As prior work \cite{HongTechRpt} has drawn a strong distinction between computational complexity and practical efficiency (with particular attention to Renegar \cite{DBLP:journals/jsc/Renegar92b}), these complexity bounds will not necessarily translate into immediate practical efficiency.
However, a followup work \cite{DBLP:journals/cj/HeintzRS93} argued for the potential of algorithms with strong theoretical complexity bounds to realize efficiency on fragments of real arithmetic, and these algorithms remain influential.

Our prior work \cite{BKR} verified the \emph{univariate} case of BKR in Isabelle/HOL; we argue there that BKR is likely more amenable to formalization than CAD, and potentially complementary to CAD.
We now extend this development \cite{BKR_AFP,BKR} into a multivariate QE algorithm.
Our multivariate algorithm is something of a hybrid: it is a mixture of Tarski's original QE algorithm \cite{Tarski} and BKR \cite{DBLP:journals/jcss/Ben-OrKR86}, with insights from Renegar \cite{DBLP:journals/jsc/Renegar92b}.
It currently does not exploit \textit{all} of the reduction from BKR, which limits its efficiency.
Thus, like Cohen and Mahboubi \cite{AssiaQE}, we view our contribution as being primarily theoretical \textit{from the perspective of efficiency}.
However, we also view our algorithm as being a significant stepping stone towards the BKR algorithm and, eventually, its variants.
In particular, it would be of considerable interest to verify a method that more closely realizes the parallel complexity bounds of Renegar \cite{DBLP:journals/jsc/Renegar92b}.
Such a method will naturally take time to develop, and will likely only be realized in stages.

\textit{Contributions.} (1) Our work is the first complete multivariate QE algorithm formalized in Isabelle/HOL.
(2) To our knowledge, it is the first formalized multivariate QE algorithm to include insights from BKR, and it is a first step towards a less complex verified algorithm (e.g. in the style of Renegar \cite{DBLP:journals/jsc/Renegar92b}), which could ideally complement an eventual formalized algorithm based on CAD.
(3) Because much of the source material is either sparsely written (e.g. \cite{DBLP:journals/jcss/Ben-OrKR86}) or highly mathematical (e.g. \cite{algRAG,DBLP:journals/jsc/Renegar92b}), it was not a priori obvious what the formalized algorithm should look like (this formalization barrier is discussed in \rref{sec:Difficult}).
The rigorous nature of verification forced us to clearly identify the essential building blocks of the algorithm: In our formalization, \emph{all} definitions are mathematically precise and verifiable, and \emph{all} their correctness properties are identified and proved.

The formalization is approximately 8500 lines of code and is available on the Archive of Formal Proofs (AFP) \cite{Multiv_BKR_AFP}.
It includes various advances to Isabelle/HOL's existing libraries, particularly the library for multivariate polynomials, which could help pave the way for future multivariate QE algorithms in Isabelle/HOL.

\section{Quantifier Elimination}\label{sec:QE}
Our QE algorithm works by eliminating one quantifier at a time.
Hence, if we have polynomials in $n + 1$ variables, we can consider them as univariate polynomials in a variable of interest with coefficient polynomials in $n$ variables. 
For example, if $x$ is our variable of interest, then we can treat $3xyz^2 + 6x^2wv + 5xy + 1$ as the following polynomial in $x$: $(6wv)x^2 + (3yz^2 + 5y)x + 1$.
For clarity, and WLOG, we assume throughout this section that our variable of interest is $x$.

The key component of both multivariate and univariate BKR is a \textit{sign-determination algorithm} which is concerned with finding all \textit{consistent sign assignments} to a set of polynomials $\{q_1, \dots, q_k\}$.
A \textit{sign assignment} is a mapping that assigns each polynomial to a \textit{sign}, i.e. positive, zero, or negative (represented by $1, 0,$ and $-1$).
A sign assignment is called \textit{consistent} if it is actually realized at some real point.

At the heart of the sign-determination algorithm that we formalize is a \textit{matrix equation} that is capable of storing sign information for a set of polynomials in variables $x, y_1, \dots, y_n$, under a set of assumptions on polynomials in $y_1, \dots, y_n$.
Our overall quantifier elimination algorithm takes a formula and identifies the polynomials that occur in the formula.
It then generates a number of matrix equations, each of which captures some sign information for the polynomials, subject to some list of assumptions.
Collectively, it is important that the generated matrix equations have exhaustive assumptions---in the sense that for every possible set of assumptions, there is at least one corresponding matrix equation.
We call sets of assumptions \textit{branches}.
Branches are refined throughout the construction with additional assumptions until each multivariate matrix equation has assumptions that generate a unique matrix equation.
Initial branches, which are not fully refined, may still have multiple associated matrix equations.

WLOG, we assume that we are eliminating a $\forall$ quantifier (because $\exists$ quantifiers can be transformed into $\forall$ quantifiers with appropriate negations). 
We do some initial branching (this is needed to guide the computations of the matrix equations), and for each branch, we check whether \textit{all} of the associated matrix equations describe a sign assignment on our polynomials that satisfies the original formula.
We filter our initial branches to pick out the ones that satisfy this property.
Finally, we return a disjunction of all assumptions of the initial branches in this filtered list.

\begin{figure}[t]
  \centering
  \includegraphics[width=\linewidth]{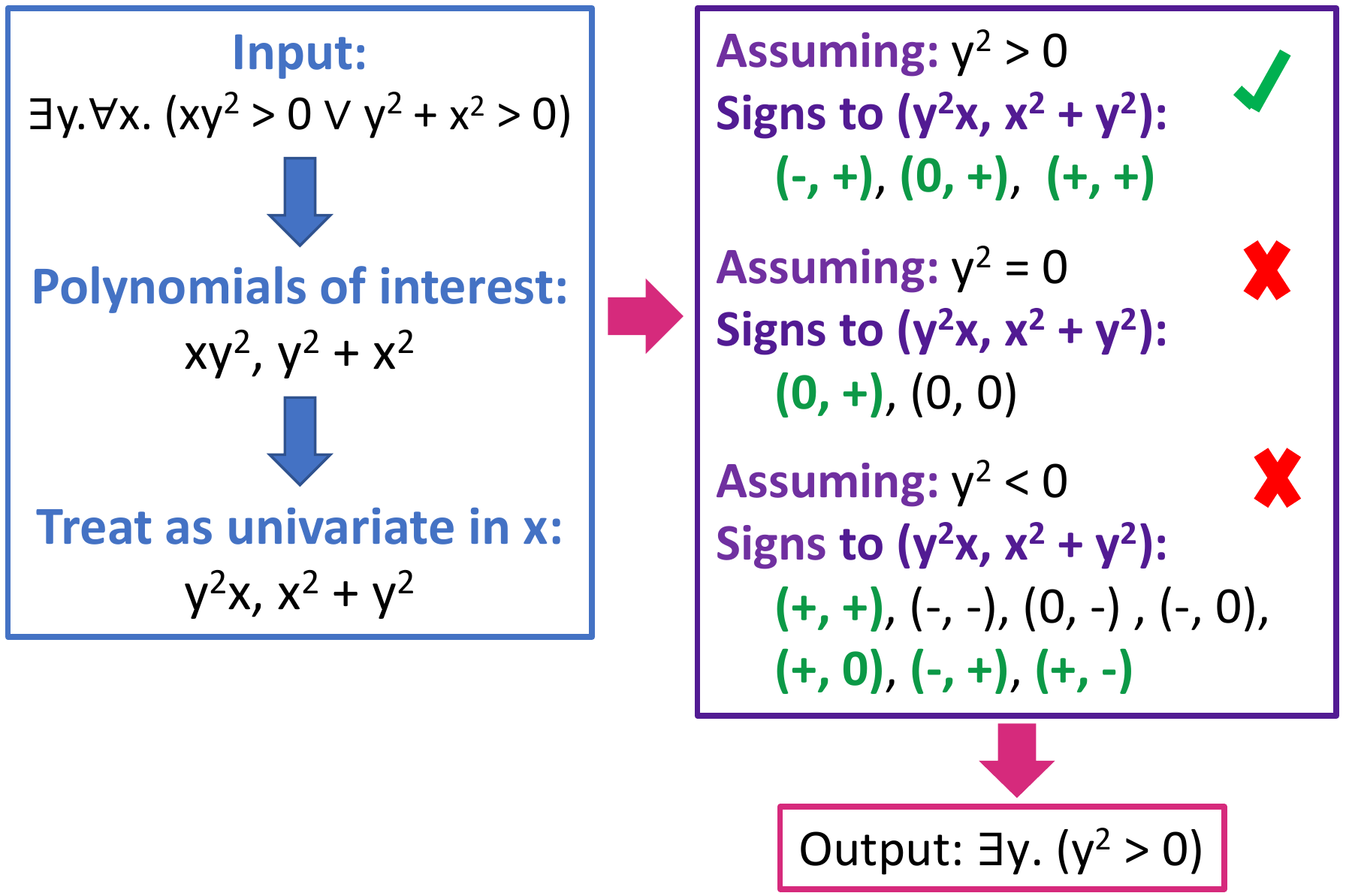}
  \caption{A visual overview of the QE algorithm.}
  \Description{A flowchart style example which transforms an input quantifier elimination problem (described in the text) into sets of consistent sign assignments subject to different assumptions and then finally into a result.}
  \label{fig:QEOverview}
\end{figure}

\rref{fig:QEOverview} visualizes how this QE algorithm works on an example.
We begin with formula $\exists y. \forall x. (xy^2>0 \lor y^2+x^2>0),$ where our focus is on eliminating the $\forall x$ quantifier.
We first identify the polynomials of interest in this formula and view them as univariate polynomials in $x$ (with coefficients that are polynomials in $y$): these are $y^2x$ and $x^2+y^2$.
Next, we determine all consistent sign assignments to these polynomials of interest given all \textit{possible}\footnote{Here, we differ from the BKR algorithm, which would branch on all \textit{consistent} sign assumptions on $y^2$.
That is, we consider a branch where $y^2 < 0$, because this is a possible (but inconsistent) sign assumption: even though $y^2$ is never negative, our algorithm does not discern this when branching.} 
sign assumptions on $y^2$, where $y^2$ is significant because it is the \textit{leading coefficient} of $y^2x$ (technically our algorithm will do some additional and unnecessary branching, but for the clarity of this example we focus on the branch on $y^2$; see \rref{sec:SignDet} for a more in-depth discussion of the branching).
Internally, our algorithm performs sign determination using matrix equation constructions (but this is not pictured in the figure).
We then pick out the sign assignments that solve our original QE problem---that is, we are looking for one of our polynomials of interest, $y^2x$ or $x^2 + y^2$, to be positive.
Signs that satisfy this condition are pictured in green.
Then, we filter our branches to find the ones where \textit{every} sign assignment satisfies the original QE problem. 
This happens only in the branch where $y^2$ is assumed to be positive.
This means that $y^2 > 0$ is logically equivalent to $\forall x. (y^2x>0 \lor x^2+y^2>0)$, which means that $\exists y. \forall x. (xy^2>0 \lor y^2+x^2>0)$ is logically equivalent to $\exists y. (y^2 > 0)$, whose quantifier $\exists y$ can be eliminated further.

If our original QE question was instead $\exists y. \forall x. (xy^2~\geq~0 \lor x^2+y^2>0),$ then both the branch with assumption $y^2 > 0$ and the branch with assumption $y^2 = 0$ would satisfy our QE problem.
This means that the disjunction $y^2 > 0 \lor y^2 = 0$ is logically equivalent to $\forall x. (y^2x\geq 0 \lor x^2+y^2>0)$, and so our output in this case would be $\exists y. (y^2 > 0 \lor y^2 = 0)$.

Here it is important to note that there are many logically equivalent outputs to any given QE problem.
For example, if our original QE question were  $\forall x. ({(xy^2= 0 \land x^2+y^2=0)} \lor {(xy^2= 0 \land x^2+y^2<0)} ),$ then two possible correct outputs that are logically equivalent are $y^2 = 0$, and $y^2<0 \lor y^2=0$.
Here, $y^2 = 0$ is the simplest output.
While the output of our QE algorithm is always logically correct, it is \textit{not} guaranteed to be in the simplest form.
In particular, assumptions for branches that are inconsistent will often be included in the final disjunction, which has no impact on logical correctness, only formula complexity.

We now turn to more detailed descriptions of the sign determination procedure, the multivariate matrix equation, and the full quantifier elimination procedure.

\subsection{Sign Determination}\label{sec:SignDet}
Finding sign information for polynomials $q_1, \dots q_k$ in variables $x, y_1, \dots, y_n$ is, on the surface, a continuous problem---the most obvious way to determine the sign information would be to evaluate $(q_1, \dots, q_k)$ on $\mathbb{R}^k$, which is clearly not computationally viable.
To account for this, BKR and Renegar reduce the sign-determination problem to a problem with the following format: find sign information for $q_1, \dots, q_k$ \emph{at the roots} of some cleverly chosen polynomial $p$.
This problem is clearly computationally viable for \emph{univariate} polynomials, because polynomials in one variable only have finitely many roots.
It is a (non-obvious) key insight that it is also computationally viable for \emph{multivariate} polynomials \cite{DBLP:journals/jcss/Ben-OrKR86, DBLP:journals/jsc/Renegar92b}.
Intuitively, the output of the univariate algorithm only depends on the \emph{signs} of the real polynomial coefficients and not on the actual \emph{values} of those coefficients.
Thus, the algorithm lifts to the multivariate case by making \emph{sign assumptions} on (multivariate) polynomial coefficients in variables $y_1, \dots, y_n$.

In our multivariate setting, $p = (\prod q_i) \cdot \frac{\partial}{\partial x}(\prod q_i)$ is chosen for $p$.
To see what makes this particular polynomial useful, consider some valuation $\nu$ on $y_1, \dots, y_n$ (i.e., some assignment of $y_1, \dots, y_n$ to real values).
Let $\nu(f)$ denote the evaluation of polynomial $f$ in valuation $\nu$; note that $\nu(f)$ is univariate in $x$.
Now, the roots of $\nu(p) = (\prod \nu(q_i)) \cdot \frac{\partial}{\partial x}(\prod \nu(q_i)) = (\prod \nu(q_i)) \cdot \frac{d}{dx}(\prod \nu(q_i))$ contain all of the roots of the $\nu(q_i)$'s (since each $\nu(q_i)$ divides $\nu(p)$), as well as sample points from intervals between the roots (by Rolle's theorem \cite{BKR}).
Because these intervals are \textit{sign-invariant}---that is, no $\nu(q_i)$ changes sign in any of these intervals, since no $\nu(q_i)$ can change sign without passing through a root---sign information at a single point within any of these intervals is \textit{representative} of sign information for the entire interval.
So, we see that the only intervals which the roots of $\nu(p)$ do not adequately cover are the extreme intervals---the leftmost and rightmost, which lie beyond any of the roots of $\nu(p)$---for which sign information can be computed with a limit calculation on the $\nu(q_i)$'s.\footnote{In the formalization of the univariate case \cite{BKR}, the polynomial $p$ was chosen so as to directly sample from these intervals by using the Cauchy root bound, a mathematical quantity that bounds the roots of a set of univariate polynomials.
This followed BKR's original work \cite{DBLP:journals/jcss/Ben-OrKR86}.
However, since the Cauchy root bound is for univariate polynomials only, we must work instead with limit computations as Renegar does \cite{DBLP:journals/jsc/Renegar92b}.}
So, this polynomial $p$ allows a natural lifting from the univariate QE algorithm to the multivariate case, but the correctness justification needs an extensive covering of the influence of all possibilities for valuation $\nu$.

\begin{figure}[t]
  \centering
  \includegraphics[width=\linewidth]{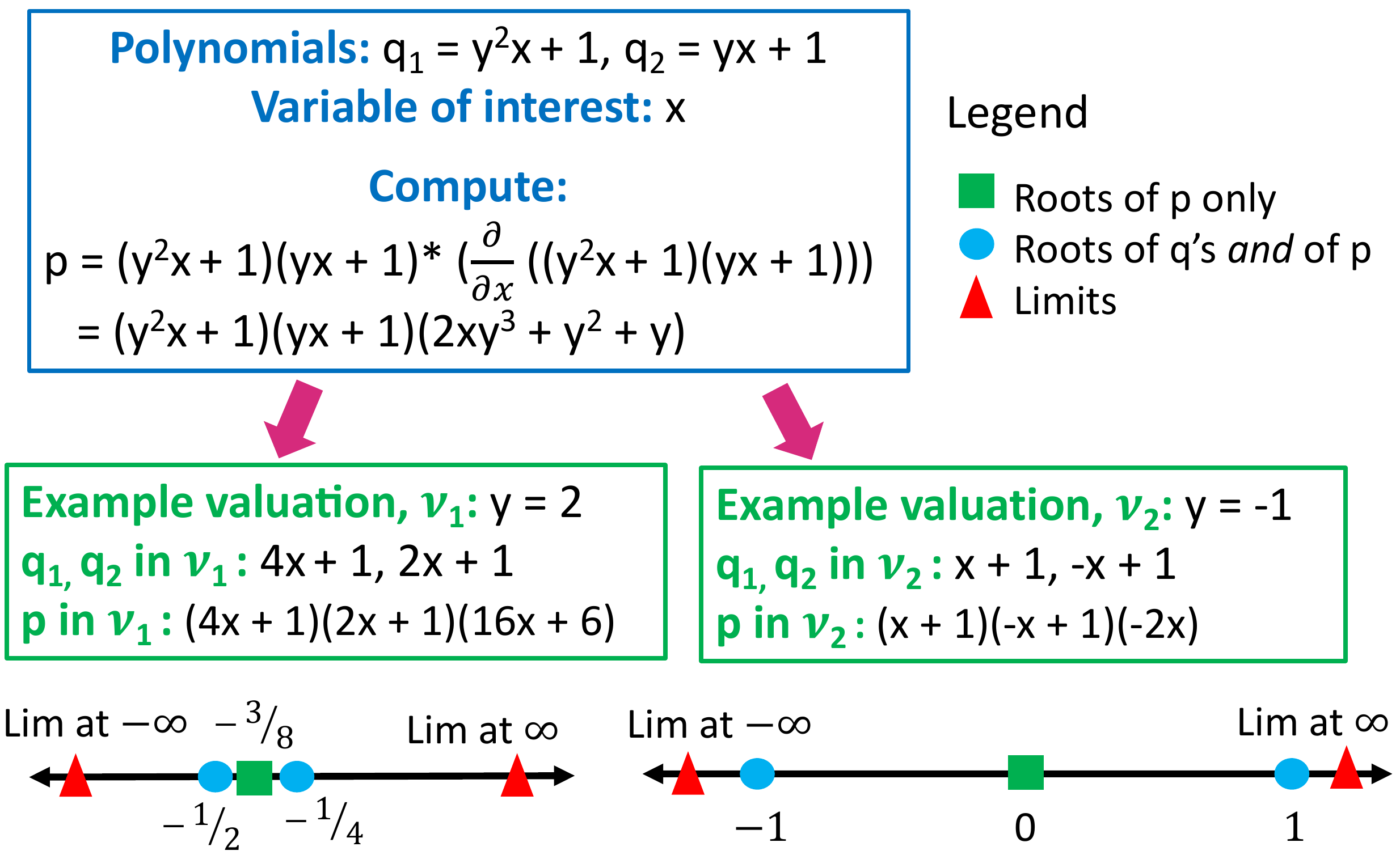}
  \caption{An example of sign determination.}
  \Description{A flowchart style example with two input polynomials, $q_1 = y^2x + 1$ and $q_2 = yx + 1$, in variable of interest $x$. From this input, the polynomial $p = (y^2x + 1)(yx + 1)(2xy^3 + y^2 + y)$ is computed and two valuations are considered.  In the first valuation, $y = 2$, so $q_1 = 4x+1$, $q_2 = 2x+1$, and $p = (4x+1)(2x+1)(16x+6)$.  It is seen that the roots of $q_1$ and $q_2$ in this valuation are $-0.5$ and $-0.25$, whereas the roots of $p$ are $-0.5$, $-0.375$, and $-0.25$. In the second valuation, $y = -1$, so $q_1 = x+1$, $q_2 = -x+1$, and $p = (x+1)(-x+1)(-2x)$.  It is seen that the roots of $q_1$ and $q_2$ in this valuation are 1 and -1, whereas the roots of $p$ are $-1, 0$, and $1$.  Thus in both valuations, taking the roots of $p$ plus the limit points covers sign information for the entire number line.}
  \label{fig:SignDet}
\end{figure}

This is visualized in \rref{fig:SignDet}.
Here, we have polynomials $q_1 = y^2x + 1$ and $q_2 = yx + 1$, so $p = (y^2x + 1)(yx + 1)(2xy^3 + y^2 + y)$.
For the purposes of illustration, we consider two sample valuations: in $\nu_1$, we set $y = 2$, and in $\nu_2$, we set $y = -1$.
As depicted, in both valuations, to find sign information for $q_1$ and $q_2$, it suffices to find sign information for $q_1$ and $q_2$ at the roots of $p$ and the limit points.

We formalize this procedure for sign determination in the \isa{sign\_determination} function.
The first input to this function is a list of polynomials \isa{qs} of type \isa{rmpoly}, where \isa{rmpoly} is our abbreviation for \isa{real mpoly poly}.
Here, \isa{poly} is \Isabelle's type for univariate polynomials, \isa{mpoly} is the type for multivariate polynomials, and \isa{real} is the type for real numbers, so an \isa{rmpoly} is a univariate polynomial whose coefficients are real multivariate polynomials.
Say initially we have polynomials in variables $x, y_1, \dots, y_n$; then type \isa{rmpoly} arises when we treat those polynomials as being univariate in $x$ with coefficients in $y_1, \dots, y_n$.
Unlike in computer algebra, these polynomials are not restricted to have any particular representation; rather, they are elements of the free term algebra.
The next input to \isa{sign\_determination} is a list of initial assumptions of type \isa{(real mpoly\ \isasymtimes\ rat)\ list}, which we abbreviate as \isa{assumps}.
Here, \isa{rat} is \Isabelle's type for rational numbers, and so each assumption in the list pairs a real multivariate polynomial with an associated rational number that indicates a sign condition on the polynomial (0, 1, or -1).
This type is useful in specifying any known sign information on polynomials in $y_1, \dots, y_n$.
The output of \isa{sign\_determination} is a list of pairs of assumptions and associated sign assignments to \isa{qs}.
Each sign assignment has type \isa{rat\ list}.\footnote{Technically, we could use \isa{int\ list} for sign assignments, since each member of the sign assignment list is $1$, $0$, or $-1$, but as noted elsewhere \cite{BKR}, it is easier to work with \isa{rat\ list} in the matrix equation construction.}
The assumptions have type \isa{assumps} (for the same reason as before), and as each assumption may have multiple associated sign assignments, each assumption is paired with a \textit{list} of associated sign assignments, as demonstrated by the \isa{assumps\ \isasymtimes\ (rat\ list\ list)} type.
The output, of type \isa{(assumps \isasymtimes\ (rat\ list\ list))\ list}, contains an exhaustive set of assumptions (in order to capture \textit{all} consistent sign assignments for the $q_i$'s).

\begin{isabelle}
\signdetermination
\end{isabelle}

Here, the \isa{lc\_assump\_generation\_list} function generates an exhaustive list of \textit{possible} branches, \isa{branches}, that contain assumptions on the signs of the leading coefficients of the input polynomials \isa{qs}.
An important subtlety is that the leading coefficient of the polynomial $q_i$ may be different in different branches.
For example, the leading coefficient of $(y+1)x^2 + yx + 2$ is $y+1$ in a branch where $y+1$ is assumed to be nonzero, $y$ in a branch where $y+1$ is zero and $y$ is assumed to be nonzero, and $2$ in a branch where both $y+1$ and $y$ are assumed to be zero.
To best account for this subtlety, each element of \isa{branches} contains both the generated assumptions (which determine the branch) \textit{and} a list of polynomials which contains a simplified version of the \isa{qs}: to be precise, $q_i = c_1x^{d_1} + \cdots + c_mx^{d_m}$ simplifies to $c_jx^{d_j} + \cdots + c_mx^{d_m}$ iff $c_1, \dots, c_{j-1}$ are all assumed to be zero and $c_j$ is assumed to be nonzero.
For example, given a list of input polynomials $[(y+1)x^2 + yx + 2, y^2 + (y+1)x^5]$, an element of \isa{branches} could be: $([(y + 1, 0), (y, 1), (y^2, 1)] , [yx + 2, y^2 + (y+1)x^5])$.
The list of assumptions $[(y + 1, 0), (y, 1),  (y^2, 1)]$ specifies that, in this branch, $y + 1$ is assumed to be 0 and $y$ and $y^2$ are assumed to be positive.
Under these assumptions, $(y+1)x^2 + yx + 2$ simplifies to $yx + 2$ and $y^2 + (y+1)x^5$ simplifies to $y^2 + (y+1)x^5$ (as the purpose of the simplification is to determine the leading coefficient, it is not mission critical to fully simplify $y^2 + (y+1)x^5$ to $y^2$, and our code is not optimized to do so).

Currently, \isa{lc\_assump\_generation\_list} naively generates branches by branching on \textit{all possible} sign assignments to the leading coefficients, rather than on all \textit{consistent} ones as BKR would. 
Thus, branches with inconsistent assumptions can be generated: for example, the branch $([(y + 1, 0), (y, 1), (y^2, -1)],$ $[yx + 2, y^2 + (y+1)x^5])$ could be generated by the function \isa{lc\_assump\_generation} despite its inconsistent assumptions ($y^2$ is assumed to be negative).
Additionally, although \isa{lc\_assump\_generation\_list}  takes an input list of assumptions, \isa{assumps}, as an argument, it does not enforce consistency of the output branches with \isa{assumps}; however, before splitting on the sign of a polynomial $f$, it will check whether \isa{assumps} already contains sign information for $f$.

Branching on the signs of the leading coefficients of the \isa{qs} provides important information for two reasons: First, because these signs are relevant for the matrix equation computation (\rref{sec:MatEq}); and second, because knowing the sign of the first non-zero leading coefficient for every $q_i$ allows us to easily compute the signs at the limit points.\footnote{The sign of $q_i$ at $\infty$ equals the sign of its leading coefficient, whereas the sign of $q_i$ at $-\infty$ is the sign of its leading coefficient multiplied by $(-1)^{\deg q_i}$, where ${\deg q_i}$ is the degree of $q_i$.}

The \isa{sign\_determination} function maps over \isa{branches}, and for each computes the polynomial $p = (\prod q_i) \cdot \frac{\partial}{\partial x}(\prod q_i)$, stored in \isa{poly\_p\_branch} (cross reference \rref{fig:SignDet}).
Although it would suffice to compute $p$ beforehand, and then simplify it appropriately on each branch given the associated assumptions (for example, in a branch where $y = 0$, $q_1 = y^2x + 1$, and $q_2 = yx + 1$, the polynomial $p = (y^2x + 1)(yx + 1)(2xy^3 + y^2 + y)$ simplifies to $p = 0$), it is more direct to compute $p$ in each branch.\footnote{Our polynomials do not have any fixed representation, and equality checking is a potentially costly operation.
Further, even if two polynomials are not identically equivalent, they may be so under a branch's assumptions (for example, $y^2 + y + 1$ is equivalent to $y^2$ if $y + 1$ is assumed to be 0).}
That is, given $q_1 = y^2x + 1$, and $q_2 = yx + 1$, if in a given branch we know that $y = 0$, we also know that the leading coefficient of $q_1$ is 1 and the leading coefficient of $q_2$ is 1, which means that $q_1 = 1$ and $q_2 = 1$, and so $p = (1\cdot 1)\cdot (\frac{\partial}{\partial x} (1 \cdot 1)) = 0$.

Next, for each branch, \isa{sign\_determination} performs a calculation (formalized in our \isa{limit\_points\_on\_branch} function) to find the signs of \isa{qs} at $\infty$ and $-\infty$.
These are stored in \isa{pos\_limit\_branch} and \isa{neg\_limit\_branch}, respectively.

Then, it makes a call to our \isa{calculate\_data\_assumps\_M} function (discussed in \rref{sec:MatEq}) to calculate a list of matrix equations for each branch, each of which stores sign information under some assumptions (assumptions in our formalization only accumulate, so the output assumptions contain the original branch's assumptions).
It pulls out the assumptions and sign conditions from the matrix equations with the \isa{extract\_signs} function, which returns a list of type \isa{(assumps\ \isasymtimes\ rat\ list\ list)\ list}.
This list is stored in \isa{mat\_eq\_signs\_on\_branch}.

Finally, the positive and negative limit sign conditions \isa{pos\_limit\_branch} and \isa{neg\_limit\_branch} are preprended to each list of sign conditions calculated with the matrix equations (the \isa{\isacharhash} operator in \Isabelle prepends an element to a list), and the resulting list of assumptions and associated sign conditions is returned.

It is now time to discuss the matrix equation.

\subsection{The Multivariate Matrix Equation}\label{sec:MatEq}
The multivariate matrix equation, like the univariate matrix equation, is concerned with finding sign information for a set of polynomials $q_1, \dots, q_n$ at the roots of an auxiliary polynomial $p$.
One advantage of formalizing a multivariate QE algorithm based on BKR and Tarski is that the construction of the multivariate matrix equation is very similar to the construction of the univariate matrix equation.

Thus, to understand the multivariate matrix equation, we first need to consider the construction of the univariate matrix equation.
At its core, the univariate matrix equation relies on computing \textit{Tarski queries}, so we start there.

\subsubsection{Computing Multivariate Tarski Queries}\label{sec:TQ}
Tarski queries are defined as follows:

\begin{definition}\cite{BKR}
Given \textit{univariate} polynomials $p, q$ with $p \neq 0$, the \textit{Tarski query} $N(p, q)$ is: 
\begin{align*}
    N(p, q) =&\ \#\{ x \in \mathbb{R} ~|~ p(x) =0, q(x) > 0 \}\ - \\
   &\ \#\{ x \in \mathbb{R} ~|~ p(x) =0, q(x) < 0\}.
\end{align*}
\end{definition}

These Tarski queries can be computed from the Euclidean remainder sequence that starts with $p$ and $p'q$: 

\begin{proposition}\label{prop:ST} (Sturm-Tarski Theorem)
Let $p \neq 0$ and $q$ be real \emph{univariate} polynomials. Let $p_1 = p$, $p_2 = p'q$, $p_3, \dots, p_k$ be the Euclidean remainder sequence of $p$ and $p'q$, where
$$p_i = c_ip_{i+1} - p_{i+2},$$
for $c_i\in\mathbb{R}[x]$ and where $\deg(p_{i+2})<\deg(p_{i+1})$.
Let $a_i$ be the leading coefficient of $p_i$ and let $d_i := \deg(p_i)$.
Let $S^+(p, q)$ denote the number of sign changes in the sequence $a_1, \dots, a_k$, and let
$S^-(p, q)$ denote the number of sign changes in the sequence $(-1)^{d_1}a_1 \dots, (-1)^{d_k} a_k$.
Then $N(p, q) = S^-(p, q) - S^+(p, q)$.
\end{proposition}

This result is from the literature \cite[Prop. 8.1] {DBLP:journals/jsc/Renegar92b} (with an unnecessary assumption removed that is not included in other references \cite{algRAG} or in Isabelle's existing formalization \cite{Sturm_Tarski-AFP} of the Sturm-Tarski theorem).
Critically, in the Sturm-Tarski theorem, it is not the values of $a_1, \dots, a_k$ that matter; rather, it is the signs that matter; this is what enables the multivariate generalization \cite{DBLP:journals/jcss/Ben-OrKR86}.

Consider polynomials $p \neq 0$ and $q$ in $x$ with polynomial coefficients in $y_1, \dots, y_n$ (i.e.,  $p, q \in \mathbb{R}[y_1, \dots, y_n][x]$).
Then, we can form Euclidean remainder sequences of $p$ and $p'q$ with respect to $x$.
The Euclidean remainder sequence is no longer unique---instead, there are multiple sequences, each depending on the signs of the coefficients of $p$ and $q$ (as coefficients that are polynomials can have different signs at different points).
Once we fix a sequence and find the leading coefficients, we need to consider (by branching) \textit{all} possible sign assignments to those coefficients,\footnote{Full BKR would consider all consistent sign assignments instead. This makes the algorithm highly recursive, which adds a considerable layer of difficulty to its verification.} and output a list of Tarski queries and the assumptions they are subject to.

For example, if we take polynomials $p = y^2x + 1$ and $q = yx + 1$, then if $y^2 = 0$, then $y = 0$ so $p = q = 1$, and the Euclidean remainder sequence is just $1$, and $N(p, q) = 0$.\footnote{Technically, our formalization would do more branching than this for two reasons: First, it will branch on $y^2 = 0$, $y^2 > 0$, and (unnecessarily) $y^2 < 0$; and second, because it will not determine that $y^2 = 0$ implies $y = 0$---and so it will not know that $q = 1$ whenever $y^2 = 0$.}
However, if $y \neq 0$, then our Euclidean remainder sequence is $y^2x + 1, y^3x + y^2, -(1 - y)$, where we have calculated $y^2x + 1 = \frac{1}{y}\cdot(y^3x + y^2) + (1 - y)$, using assumption $y \neq 0$ for $\frac{1}{y}$.

\begin{figure}[t]
  \centering
  \includegraphics[width=\linewidth]{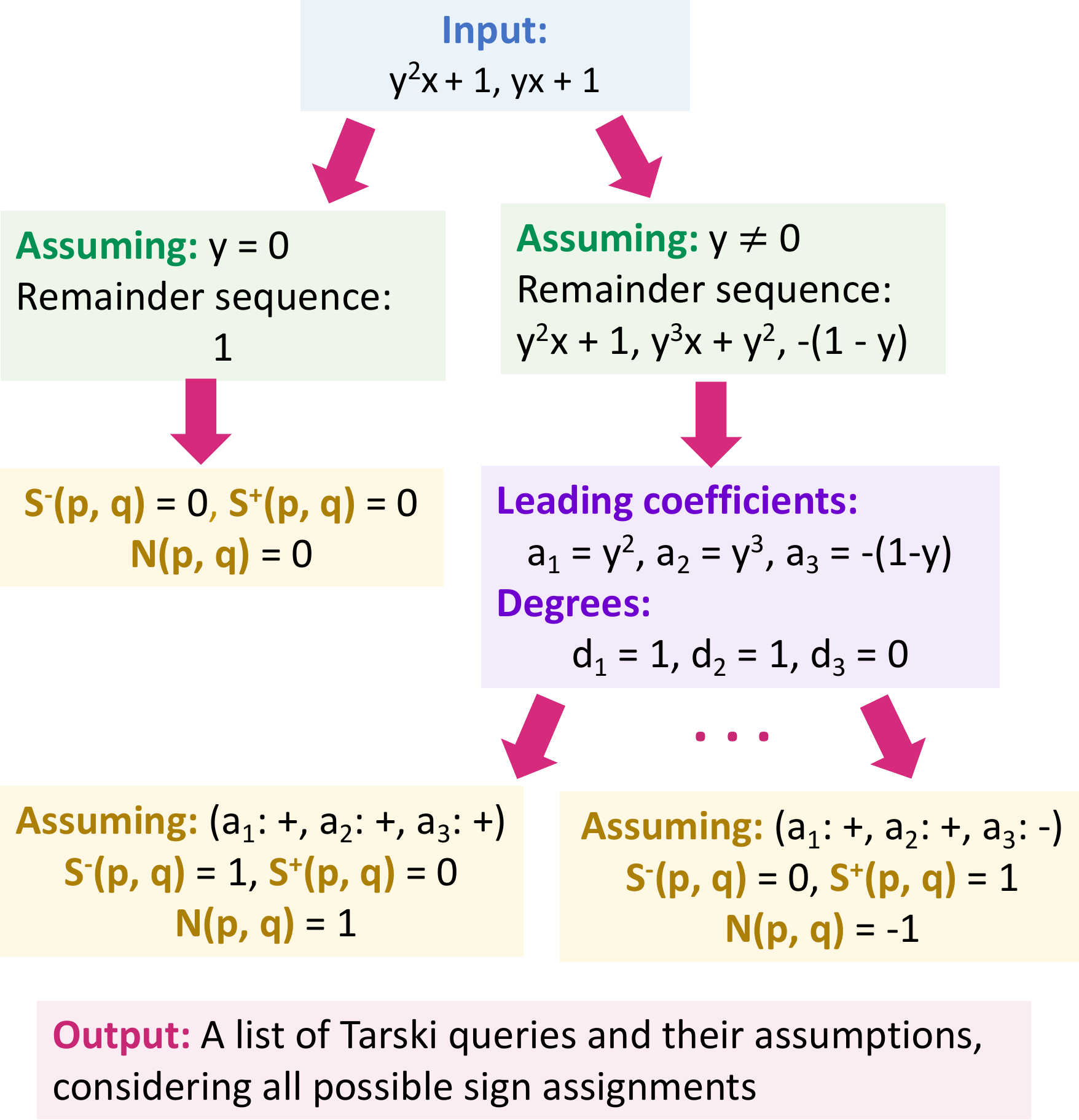}
  \caption{Computing Tarski queries for $p = y^2x + 1$, $q = yx + 1$.}
  \Description{A flowchart style example (described in the text) which visualizes computing a multivariate Tarski query.}
  \label{fig:MultivTQ}
\end{figure}

Now, continuing the computation of $N(y^2x + 1, yx + 1)$, we find that the leading coefficients of our Euclidean remainder sequence (assuming $y \neq 0$) are $y^2, y^3,$ and $-(1-y)$.
Next, we consider the possible sign assignments to $y^2, y^3$, and $-(1 -y)$. 
For example, $(+, +, -)$ is one such sign assignment.
So, we have Tarski query $N(p, q) = S^-(p, q) - S^+(p, q) = 0 - 1 = -1$ under the assumptions that: $y \neq 0$, $y^2 > 0$, $y^3 > 0$, and $-(1 - y) < 0$.
Our output for $N(y^2x + 1, yx + 1)$ would be a list of all the Tarski queries under all possible assumptions.
This computation is visualized in \rref{fig:MultivTQ} (where, for purposes of space, only two output branches are shown explicitly).

Note that Euclidean remainder sequences for multivariate polynomials sometimes contain fractions.
While we could have chosen to work with Euclidean remainder sequences in a \textit{fraction field}, this would require complicated type switching in the formalization.
Instead, we use \textit{pseudo-remainder sequences} for multivariate polynomials.
Pseudo-remainder sequences are essentially Euclidean remainder sequences for polynomials, but normalized so as not to contain fractions (ours are additionally normalized so as not to affect the result of the Sturm-Tarski computation \cite{li2019deciding}).
We develop pseudo-remainder sequences for multivariate polynomials of type \isa{rmpoly} (currently, our formalization naively branches on the signs of the leading coefficients of the relevant polynomials).
Here, we benefit from prior work: The Sturm-Tarski theorem was formalized in Isabelle/HOL by Wenda Li \cite{Sturm_Tarski-AFP}; Li and Paulson later extended this to \textit{bivariate} polynomials \cite{li2016modular} using pseudo-remainder sequences, and Li, Passmore, and Paulson also developed univariate Tarski queries with pseudo-remainder sequences \cite{li2019deciding}.

\begin{remark}
For self-containedness, we briefly describe pseudo-remainder sequences.
\emph{Polynomial pseudo-quotients} (pquo) and \emph{pseudo-remainders} (prem) satisfy this property \cite{li2019deciding,DBLP:conf/cade/MouraP13}:
\[(\text{lead\_coeff }q)^{(1 + \text{deg }p - \text{deg }q)}p = \text{pquo}(p, q)\cdot q+\text{prem}(p, q),\]
where $\text{deg prem}(p, q) < \text{deg }q$ or $q = 0$.
For example, when considering polynomials $p = yx^2 + 1$ and $q = y^3x + 1$ as univariate polynomials in $x$, then $\text{pquo}(p, q) = y^4x-y$ and $\text{prem}(p, q) = y^6 + y$, as $(y^3)^2 p = (y^4x - y)q + (y^6 + y)$ and $\text{deg}(y^6 + y) = 0 < \text{deg }q = 1$.
Notice how there are no fractions in pquo or prem, unlike the fractions in the usual Euclidean remainder sequence (assuming $y \neq 0$ for well-definedness).

We use \emph{signed} pseudo-remainder sequences, where $p_1 = p$, $p_2 = p'q$, and $p_3, \dots, p_k$ satisfy the following equation for a special choice of coefficients $s_i$, explained below:
\[ p_{i+2} = s_i \cdot \text{prem}(p_{i}, p_{i+1}) \]
This sequence is normalized so that, in any valuation, the number of sign changes in the evaluated pseudo-remainder sequence is the same as in the Euclidean remainder sequence for the evaluated polynomials, so that the result of the Sturm-Tarski computation is unaffected by the normalization.
For this, we follow the style of \cite{li2019deciding} and normalize as follows: if $(1 + \text{deg }p_{i} - \text{deg }p_{i+1})$ is even, we multiply $\text{prem}(p_i, p_{i+1})$ by $s_i = -1$; else, by $s_i = -\text{lead\_coeff }p_{i+1}$.
To understand this intuitively, note that the pseudo-remainder $\text{prem}(p, q)$ effectively normalizes by $(\text{lead\_coeff }q)^{(1 + \text{deg }p - \text{deg }q)}$.
Then, note that remainder sequences in the Sturm-Tarski theorem always negate \text{prem} (cross-reference \rref{prop:ST}).
So, if $(1 + \text{deg }p - \text{deg }q)$ is even, we have not changed the sign of prem and we need only negate it.
However, if $(1 + \text{deg }p - \text{deg }q)$ is odd, we have potentially changed the sign of prem---depending on the sign of $(\text{lead\_coeff }q)$---so we not only negate prem but also multiply it by $(\text{lead\_coeff }q)$.
\end{remark}

Since QE is concerned with sign information for multiple polynomials simultaneously, it is useful to generalize the notion of Tarski queries to \textit{sets} of polynomials \cite{BKR} as follows:

\begin{definition}
Given a polynomial $p$ and a list of polynomials $q_1, \dots, q_n$, let $I$ and $J$ be subsets of $\{1, \dots, n\}$.
Then, the Tarski query $N(I, J)$ with respect to $p$ is 
\begin{align*}
    N(&I, J) = N(p^2 + \left(\Sigma_{i \in I} q_i^2\right), \Pi_{j \in J}~q_j) =\\
    &\#\{ x \in \mathbb{R} ~|~ p(x) =0, \forall i \in I. ~q_i(x) = 0, \Pi_{j \in J}~q_j(x) > 0 \}\ - \\
   &\ \#\{ x \in \mathbb{R} ~|~ p(x) =0, \forall i \in I. ~q_i(x) = 0, \Pi_{j \in J}~q_j(x) < 0\}.
\end{align*}
\end{definition}

The matrix equation determines the signs of $q_1,$ $\dots, q_n$ at the zeros of $p$ by computing $N(I, J)$ for a representative set of combinations of subsets $I, J$ of $q_1, \dots, q_n$ (see \rref{sec:usingtq}).

There are two key lemmas that we prove about multivariate Tarski queries.
The first is a soundness lemma showing that the resulting multivariate Tarski queries agree, on every point satisfying the associated assumptions, with what the univariate Tarski query would have been:

\begin{isabelle}
\multivTQone
\end{isabelle}
Here, the \isa{construct\_NofI\_M} function constructs a list of multivariate Tarski queries and the assumptions they are subject to.
As input, it takes a polynomial \isa{p}, an initial set of assumptions \isa{acc}, and two lists of polynomials \isa{I} and \isa{J}.
Both \isa{p} and all of the polynomials in \isa{I} and \isa{J} have type \isa{rmpoly}, i.e. they are univariate polynomials in $x$ with polynomial coefficients in some variables $y_1, \dots, y_n$.
The \isa{inset} assumption assumes that we have some particular Tarski query \isa{tarski\_query} that is subject to the assumptions \isa{assumps}, which are assumptions on polynomials in $y_1, \dots, y_n$.
Now, the \isa{construct\_NofI\_R} function is the function to compute univariate Tarski queries from our prior work \cite{BKR}, so the conclusion of the lemma is that \isa{tarski\_query} is exactly the (unique) univariate Tarski query that would be computed from evaluating \isa{p} and all of the polynomials in \isa{I}, \isa{J} on \isa{val} (using the \isa{eval\_mpoly\_poly} and \isa{eval\_mpoly\_poly\_list} functions), where \isa{val} is any  assignment of real values to $y_1, \dots y_n$ where the assumptions \isa{assumps} are realized.

The second key lemma is a completeness result:
\begin{isabelle}
\multivTQtwo
\end{isabelle}
Here, this shows that if initial assumptions \isa{init\_assumps} are satisfied by valuation \isa{val}, then there is some resulting assumptions and Tarski query pair \isa{(assumps, tq)} where all final assumptions \isa{assumps} are satisfied by \isa{val}.

Together, these two lemmas give a strong result: the soundness lemma shows that the multivariate results coincide with univariate results in all projections meeting the final assumptions, and the completeness lemma shows that for any projection meeting the initial assumptions, there is some corresponding Tarski query whose associated (final) assumptions are met by the projection.
Or, on a more intuitive level, the completeness lemma shows that our function to compute multivariate Tarski queries generates useful output whenever it is given useful input, and the soundness lemma shows that useful output has the desired mathematical meaning.

\subsubsection{Using Multivariate Tarski Queries}\label{sec:usingtq}
The matrix equation connects a vector of information about \textit{possible sign assignments} for a set of multivariate polynomials---i.e., sign assignments that are not necessarily consistent---on the LHS, to a vector of multivariate Tarski queries on the RHS.

The univariate matrix equation is defined as follows, where we closely follow the definition of the univariate matrix equation in our earlier work \cite{BKR}, but adapted to our purposes:\footnote{The univariate BKR paper \cite{BKR} follows the matrix equation developed in Ben-Or, Kozen, and Reif's original paper \cite{DBLP:journals/jcss/Ben-OrKR86}, where $p$ is assumed to be coprime with each $q_i$.
Because this assumption no longer makes sense for multivariate polynomials, we use the matrix equation developed by Renegar \cite{DBLP:journals/jsc/Renegar92b}.
While our prior work \cite{BKR} formalized both styles of matrix equation \cite{BKR_AFP}, only the former was discussed at length in the paper.}
\begin{definition}
Fix univariate polynomials of interest $p$ and $q_1, \dots, q_k$.
Let $\tilde{\Sigma} = \{\tilde{\sigma}_1, \dots, \tilde{\sigma}_m\}$ be a set of possible sign assignments to $q_1, \dots, q_k$, and assume $\tilde{\Sigma}$ contains all consistent sign assignments to $q_1, \dots, q_k$ at the roots of $p$.
Let $S$ be a set of pairs of subsets $(I_1, J_1),$ $\dots,$ $(I_{l}, J_{l})$ where for all $1 \leq i \leq l$, $I_i \subseteq \{1, \dots, k\}$ and $J_i \subseteq \{1, \dots, k\}$.  
Then the \emph{matrix equation} for $\tilde{\Sigma}$ and $S$ is the relationship $M \cdot w = v$ between the following three entities:
\begin{itemize}
\item $M$, the $l$-by-$m$ matrix with entries \[M_{i,j} = \left(\Pi_{\ell \in I_i} (1 - (\tilde{\sigma}_j(q_\ell))^2)\right) \cdot \left(\Pi_{\ell \in J_i} \tilde{\sigma}_j(q_\ell)\right)\in\{-1,0,1\}\] for $(I_i,J_i) \in S$ and $\tilde{\sigma}_j \in \tilde{\Sigma}$,
\item $w$, the length $m$ vector whose entries count the number of roots of $p$ where $q_1, \dots, q_k$ has sign assignment $\tilde{\sigma}$, i.e., $ w_i = \#\{x \in \mathbb{R} ~|~ p(x) = 0, \text{sgn}(q_{\ell}(x)) = \tilde{\sigma}_i(q_{\ell}) ~\text{for all}~ 1\leq \ell \leq k\}$,
\item $v$, the length $l$ vector consisting of Tarski queries for the subsets, i.e., $v_i =  N(I_i, J_i)$.
\end{itemize}
\end{definition}

Intuitively, as noted by our prior work \cite{BKR}, the meaning of a matrix equation is captured by its associated list of signs and list of (pairs of) subsets.
Both the matrix $M$ and the RHS vector $v$ are fully computable from these two lists, and $w$, which stores information about which possible sign assignments are consistent (sign assignment $\tilde{\sigma}_i$ is consistent iff $w_i$ is nonzero), is calculated as $M^{-1}\cdot v$.

For multivariate polynomials the situation is more complicated.
We can still construct a matrix equation for multivariate polynomials---the definition of the matrix $M$ is the same as it was in the univariate setting, but the righthandside vector uses our function to construct a list of Tarski queries for multivariate polynomials.
Each RHS vector---and so each matrix equation---comes with an associated list of assumptions which were generated by the multivariate Tarski queries.
So, for an input list of multivariate polynomials $p$ and $q_1, \dots, q_k$, we construct a \textit{list} of multivariate matrix equations that store sign information for these polynomials, subject to certain assumptions on polynomials in one fewer variable.

The overall construction is very similar to that in the univariate case \cite{BKR}.
It proceeds by induction on the number of $q$'s, so that the base case is for a single $q$.
Smaller matrix equations are successively combined and reduced to form the matrix equation for $q_1, \dots, q_n$.
The reduction is what differentiates the matrix equation of BKR from that of Tarski: information for inconsistent sign assignments is removed at appropriate intervals, which decreases the size of the matrix equation.
In the univariate case, the size of the matrix equation is bounded by $\#\{x.\ p(x) = 0\})^2$, where $\#\{x.\ p(x) = 0\}$ is the number of roots of the polynomial $p$.
The size of a multivariate matrix equation is bounded by the number of roots of $p$ in a valuation satisfying the associated assumptions.
As the univariate reduction step mainly involves computations on the matrix $M$, which is unchanged in the multivariate setting, it generalizes quite naturally, and so our hybrid algorithm essentially inherits reduction in the matrix equation construction, thus incorporating insights from BKR into our hybrid algorithm.

We formalize our multivariate matrix equation construction in the \isa{calculate\_data\_assumps\_M} function (cross reference \rref{sec:SignDet}), and prove the following two key lemmas:
\begin{isabelle}
\calcdatacorrect
\end{isabelle}

This first lemma connects the behavior of our multivariate matrix equation constructor function to the Renegar-style univariate matrix equation function (\isa{calculate\_data\_R}) formalized in our prior work \cite{BKR}. 
That is, on any valuation \isa{val} that satisfies the assumptions \isa{assumps}, the associated multivariate matrix equation \isa{mat\_eq}, which finds the consistent sign assignments for \isa{qs} at the zeros of some \isa{p} in the valuation \isa{val}, is equal to the univariate matrix equation that find the consistent sign assignments for \isa{eval\_qs} at the zeros of \isa{eval\_p} , where \isa{eval\_p} is \isa{p} evaluated on \isa{val} and \isa{eval\_qs} is \isa{qs} evaluated on \isa{val}.
This is a soundness lemma, since it explains that whenever our output is useful, it has the correct mathematical meaning.

\begin{isabelle}
\calcdatacomplete
\end{isabelle}

This second lemma shows that when we give logically consistent input assumptions to \isa{calculate\_data\_assumps\_M}, some output with logically consistent assumptions will be generated (i.e., useful input generates useful output).
These lemmas are analogous to those discussed for multivariate Tarski queries; taken together, they help us prove key correctness properties of our \isa{elim\_forall} method, which serves to eliminate a single universal quantifier.
We now turn to a discussion of our top-level QE methods, including \isa{elim\_forall}.

\subsection{Overall Quantifier Elimination Algorithm}\label{sec:qe}
To best explain our formalized QE algorithm, we must first touch on the framework we are working with.

We build on our prior framework \cite{scharager2021verified} that verified (in Isabelle/HOL) the virtual substitution algorithm, an efficient QE method that applies to a low-degree fragment of real arithmetic.
This prior development sets up a framework for multivariate QE (including a type for real QE problems and a function to evaluate QE problems at real-valued points); by building on this, we are ultimately able to link together our verified (complete, inefficient) QE method with verified virtual substitution \cite{scharager2021verified}, using this (incomplete but experimentally promising) QE method as a preprocessing step.

Accordingly, we work with formulas of type \isa{atom\ fm}  \cite{scharager2021verified}, which have the following grammar:
\begin{align*}
	F, G~&\bebecomes~\text{TrueF}~ \alternative \text{FalseF} \alternative (\text{Atom}(\text{Eq}\ p)) \alternative  (\text{Atom}(\text{Less}\ p))\alternative \\   &\ (\text{Atom}(\text{Leq}\ p)) \alternative (\text{Atom}(\text{Neq}\ p)) \alternative 
\text{And}\ F\ G \alternative \text{Or}\ F\ G \alternative \\ 
&\ \ \text{Neg}\ F	\alternative \text{ExQ}\ F \alternative \text{AllQ}\ F \alternative \text{ExN}\ n\ F \alternative \text{AllN}\ n\ F,
\end{align*}
where $p$ is a real polynomial and $n \in \mathbb{N}$.
Here, $(\text{Atom}(\text{Eq}\ p))$ captures the relationship $p = 0$, $(\text{Atom}(\text{Less}\ p))$ captures $p < 0$, $(\text{Atom}(\text{Leq}\ p))$ captures $p \leq 0$, and $(\text{Atom}(\text{Neq}\ p))$ captures $p \neq 0$.
Further, $\text{And}\ F\ G$ captures the logical meaning of $F \land G$,  $\text{Or}\ F\ G$ captures $F \lor G$, and $\text{Neg}\ F$ captures $\lnot F$.
Finally, $\text{ExQ}\ F$ indicates that formula $F$ is quantified by an existential quantifier, $\text{AllQ}\ F$ indicates that $F$ is quantified by a universal quantifier,  $\text{ExN}\ n\ F$ indicates that $F$ is quantified by a  block of $n$ existential quantifiers, and $\text{AllN}\ n\ F$ indicates that $F$ is quantified by a  block of $n$ universal quantifiers.

In these formulas, variables are represented with de Bruijn indices; \isa{Var\ 0} is the variable quantified by the innermost quantifier, \isa{Var\ 1} is the variable quantified by the second innermost quantifier, and so on.
We operate on quantifiers inside-out, i.e. we start with the quantifier attached to \isa{Var 0}.

Our \isa{elim\_forall} function is designed to eliminate a single $\forall$ quantifier.
It parallels the method visualized in \rref{fig:QEOverview}.
\begin{isabelle}
\elimforall
\end{isabelle}
Here, \isa{extract\_polys} finds the polynomials \isa{qs} in our formula \isa{F}, and \isa{univariate\_in\ qs\ 0} transforms our polynomials \isa{qs} to have the \isa{rmpoly} type (so that they are univariate polynomials in \isa{Var 0}, with coefficients that are multivariate polynomials in bigger variables).
The resulting list of polynomials is called \isa{univ\_qs}.
Then, in \isa{reindexed\_univ\_qs}, we transform the coefficients of every polynomial in \isa{univ\_qs} (which do not contain \isa{Var 0}) by lowering every variable index by 1.
This lowering is crucial for finding all possible signs/assumptions pairs for our multivariate polynomial coefficients (cross reference \rref{sec:SignDet}), as \isa{sign\_determination} expects polynomials in \isa{Var\ 0}.
We then retain all the sign assignments that satisfy our formula of interest, and return a disjunction of the associated assumptions.
If our original formula involved polynomials in variables \isa{Var\ 0}, \isa{Var\ 1}, \dots, \isa{Var\ n}, then, because of the transformation and reindexing, these assumptions will be polynomials in variables \isa{Var\ 0}, \dots, \isa{Var\ (n - 1)}.
Our new \isa{Var\ 0}, which was previously \isa{Var\ 1}, will correctly match to the new innermost quantifier, which was previously the second innermost quantifier, and so on.

Our top-level QE method, named \isa{qe}, heavily relies on \isa{elim\_forall} and \isa{elim\_exist} (where \isa{elim\_exist\ F} is defined as \isa{Neg\ \isacharparenleft elim\_forall\ \isacharparenleft Neg\ F\isacharparenright \isacharparenright}):
\begin{isabelle}
\qe 
\end{isabelle}

Our top-level correctness theorem says that for any assignment \isa{\isasymnu} of the free variables in \isa{F} to real numbers, our original formula \isa{F} has the same truth-value as \isa{qe\ F}; or, in other words, \isa{F} and \isa{qe\ F} are logically equivalent:
\begin{isabelle}
\qecorrect
\end{isabelle}
Here, \isa{eval} is the function formalized by Scharager \emph{et al.} \cite{scharager2021verified} to evaluate formulas of type \isa{atom\ fm} on valuations.
This function accounts for the reindexing of free variables that naturally takes place during QE.
For example, $\forall x.\ x^2y \leq 0$ is logically equivalent to $y \leq 0$, but since variables are represented with de Bruijn indices, where the innermost quantifier corresponds with \isa{Var\ 0}, $\forall x.\ x^2y \leq 0$  is represented in the \isa{atom\ fm} type as \isa{AllQ (Leq\ ((Var\ 0)\isacharcircum 2\ \isasymcdot\ Var\ 1))} whereas $y \leq 0$ is represented as \isa{Leq (Var 0)}.
In \isa{eval}, this subtlety is handled by defining, e.g., \isa{\ eval\ (AllQ\ F)\ v} as \isa{(\isasymforall\ x.\ (eval\ F\ (x\isacharhash v)))}, where \isa{x\isacharhash v} is the list with head \isa{x} and tail \isa{v}.
So, \isa{qe\_correct} shows that \isa{F} evaluated on any mapping of free variables to real numbers is equal to \isa{qe\ F} evaluated on that same mapping, which establishes that \isa{qe} is sound.

We also show that \isa{qe} fully removes quantifiers in the following lemma, where \isa{countQuantifiers} counts the number of existential or universal quantifiers in formula \isa{F}: 
\begin{isabelle}
\qeremoves
\end{isabelle}

\noindent This result shows that \isa{qe} is complete.

To our knowledge, \isa{qe} is the first sound and complete algorithm for real QE to be formalized in \Isabelle (previous work \cite{li2019deciding,DBLP:journals/jar/Nipkow10, scharager2021verified} was sound but not complete).
We now turn to some further details regarding our formalization.

\section{Formalization Details}
\Isabelle is well-suited for us; we not only benefit considerably from the well-devel\-oped libraries (including aforementioned prior work \cite{BKR,li2019deciding,scharager2021verified}), but also from \Isabelle's support for automated proof search in Sledgehammer \cite{DBLP:conf/lpar/PaulsonB10}.

However, at the same time, working in the formal setting of \Isabelle poses considerable challenges.
In this section, we begin by discussing some of those challenges, followed by some of the high-level proof techniques that helped us succeed in our formalization.
We then discuss some useful low-level details regarding our extensions to \Isabelle's multivariate polynomials library.
Finally, we discuss our code export and the performance of our algorithm.

\subsection{Challenges}\label{sec:Difficult}
Many design decisions for the functions described in \rref{sec:QE} were not initially evident.
For example, the need to consistently track assumptions and pass them in as an argument to our functions throughout the calculation of the matrix equation was initially not obvious.
At first, we wrote a function that was nearly identical to \isa{calculate\_data\_assumps\_M}, with the one major difference that we did not include \isa{assumps} as an argument to this function.
While this function was fully capable of generating a multivariate matrix equation, we soon realized we had made a major mistake when we tried to extend it into a larger QE algorithm.
After this, we were careful to always include an argument for assumptions in our functions if it could possibly be applicable, regardless of whether or not it seemed immediately relevant.

The challenge of correctly formalizing the algorithm in \Isabelle is heightened because the precision of formalization sometimes identifies details that were underspecified in the source material.
Indeed, BKR's discussion of the multivariate QE algorithm was limited to only two pages and proceeds at a very high level \cite{DBLP:journals/jcss/Ben-OrKR86}.
Renegar \cite{DBLP:journals/jsc/Renegar92b} is considerably more detailed, but is also written in the style of mathematics, which necessitates significant translation to the level of formalization.
For example, the way in which the limit point calculation should be formalized, while entirely obvious in retrospect, did not become clear to us until we fixed a method of branching---and indeed, our initial method of formalizing the limit point calculation, which was agnostic to branching, did not make it into the final code for the algorithm.
Of this calculation, Renegar writes the following, in which he uses the notation $g_i$ where we use $q_i$, and $f$ instead of $p$ \cite{DBLP:journals/jsc/Renegar92b}:
``. . . each consistent sign vector of $\{g_i\}_i$ occurs at some real zero of $f$ except, perhaps, for the sign vectors of points to the right or left of all real zeros of $\prod_i g_i$.
However, the latter two consistent sign vectors are trivially determined from the leading coefficients of the polynomials $g_i$.''
While this completely describes the mathematical use of the limit point calculations, it took some time to translate it into \Isabelle definitions and proofs.

Finally, a last challenge is that even simple details can become complex in the formalized setting of a theorem prover.
For example, working with multivariate polynomials in \Isabelle poses a challenge, as the formal setting requires rigor even for operations that are simple on paper but may become much more involved when formalized.
For example, the transformation to treat a multivariate polynomial as univariate in some variable of interest is immediate on paper, but in Isabelle/HOL it is more subtle, precisely because the type of our object is changing: $3xyz^2 + 6x^2wv + 5xy + 1$ has type \isa{real\ mpoly}, whereas $(6wv)x^2 + (3yz^2 + 5y)x + 1$ has type \isa{rmpoly} (see also \rref{sec:SignDet}).

\subsection{High Level Proof Techniques}
Though treating multivariate polynomials as univariate in some variable of interest poses low-level challenges in our formal setting, it affords significant high-level simplifications.
Many of our proofs rely on the technique of universal \textit{projection}---we assume fixed real values for all variables aside from a variable of interest, which lets us work with \textit{truly} univariate polynomials.
Projection allows us to connect functions in our multivariate construction to corresponding functions in the univariate construction from our prior work \cite{BKR}.
This works because the multivariate case of the BKR algorithm builds rather directly on the univariate case, making it amenable to formalization, as noted previously \cite{BKR}.

In consequence, each key function involved in the construction of the multivariate matrix equation requires two top-level associated lemmas.
The first is a soundness lemma which connects the behavior of the multivariate function to a corresponding univariate function \cite{BKR} through projection.
The second is a completeness lemma which establishes that data for all possible projections is captured by the function for some assumptions.
Some examples of these soundness and completeness lemmas are seen in \rref{sec:MatEq} (e.g. the soundness lemma \isa{multiv\_tarski\_query\_correct} and the completeness lemma \isa{multiv\_tarski\_query\_complete}); there are many more in the actual proof development.
This proof structure does not seek to closely mimic the (highly mathematical) proofs in the source material \cite{BKR,DBLP:journals/jsc/Renegar92b}, but rather to translate the key intuition into a shape which is amenable to formalization.

Our construction and proofs are designed to be modular, and we often rely on induction to prove key properties of helper functions.
In particular, we found it very helpful to use custom induction theorems, supplementing those automatically generated by \Isabelle.
For example, the \isa{spmods\_multiv\_aux} function shown (abridged) below computes a list of pseudo-remainder sequences for polynomials \isa{p} and \isa{q} together with corresponding sign assumptions on the leading coefficients of the polynomials in each sequence.

\begin{isabelle}
\spmodsauxmanual
\end{isabelle}

The function branches depending on whether \isa{q} is the zero polynomial, otherwise, it recurses on the (possible) signs of its leading coefficient \isa{lead\_coeff q}.
Here, \isa{assumps} specifies a list of assumed input sign conditions, which are checked for assumptions on \isa{lead\_coeff q}.
Notably, \isa{spmods\_multiv\_aux} is \emph{not} structurally recursive; its termination uses the fact that, on each recursive call, the degree of the polynomial arguments \isa{one\_less\_degree q} or \isa{mul\_pseudo\_mod p q} strictly decreases.
For such functions, \Isabelle automatically generates induction theorems, but these theorems lack the usual case-splitting support for structurally recursive functions~\cite{DBLP:conf/mkm/Wenzel06}.
The following snippet shows the \Isabelle subgoal (\verb|cases|) structure that results from applying induction with the generated theorem for \isa{spmods\_multiv\_aux}.
{\small\begin{verbatim}
// apply (induct ... spmods_multiv_aux.induct)
Proof outline with cases:
  case (1 p q assumps)
  ...
qed
\end{verbatim}}

Although \isa{spmods\_multiv\_aux.induct} can, \emph{in principle}, be used to prove the aforementioned soundness and completeness properties for \isa{spmods\_multiv\_aux}, we found the proofs tedious in practice because they lack the case structuring benefits of \Isabelle's structured proof language~\cite{DBLP:conf/mkm/Wenzel06}.
Instead, we manually prove an alternative induction theorem that mimics the branching structure of \isa{spmods\_multiv\_aux} (one base case, three branches with recursion).
As before, a snippet of the \Isabelle subgoal (cases) structure is shown below (comments illustrate the branching structure).
{\small\begin{verbatim}
// apply (induct ... spmods_multiv_aux_induct)
Proof outline with cases:
  case (Base p q assumps)
  ... // base case (q = 0)
next
  case (Rec p q assumps)
  ... // lookup_assump_aux returns None
next
  case (Lookup0 p q assumps)
  ... // lookup_assump_aux returns Some 0
next
  case (LookupN0 p q assumps r)
  ... // otherwise
qed
\end{verbatim}}

Though some manual effort is needed to state and prove \isa{spmods\_multiv\_aux\_induct}, our subsequent, repeated use of this customized induction theorem makes it well worth the initial investment.
We expect similar induction theorems to be broadly useful for structuring proofs about non-structural recursive functions, including in other proof assistants.
Indeed, manual induction theorems are also used elsewhere in the development, particularly to verify invariant properties of the helper function that underlies the branching function \isa{lc\_assump\_generation\_list} (see \rref{sec:SignDet}).

\subsection{Library Extensions}\label{sec:LibExt}
We turn to some of our key results for multivariate polynomials and the library extensions they prompted.

As seen in \rref{sec:Difficult}, we need a function to convert polynomials of type \isa{real\ mpoly} to polynomials of type \isa{real\ mpoly\ poly}.
Eberl and Thiemann formalized one such way of doing this in their \isa{mpoly\_to\_mpoly\_poly} definition \cite{Factor_Algebraic_Polynomial-AFP}.
We provide the following alternate definition, which is executable:
\begin{isabelle}
\mpolytopolyalt
\end{isabelle}
This definition applies to multivariate polynomials with coefficients in a commutative ring with unity (denoted by \isa{comm\_ring\_1}).
It relies on the \isa{isolate\_variable\_sparse} function \cite{Virtual_Substitution-AFP}, where \isa{isolate\_variable\_sparse\ p\ x\ i} finds the coefficient of \isa{x\isacharcircum i} in \isa{p}.
For each \isa{i} from 0 to the degree of \isa{x} in \isa{p}, we find this coefficient and construct a monomial of type \isa{poly} with degree \isa{i} and this coefficient.
Our final polynomial is the sum of all of these monomials.

We connect our new definition to \isa{mpoly\_to\_mpoly\_poly} in the following lemma:
\begin{isabelle}
\multivasuniv
\end{isabelle}
This enables a natural interface between Eberl and Thiemann's work \cite{Factor_Algebraic_Polynomial-AFP} and the large and powerful collection of lemmas regarding \isa{isolate\_variable\_sparse}  \cite{Virtual_Substitution-AFP}, from which we benefit in the formalization.

We benefit from Eberl and Thiemann's lemmas regarding \isa{mpoly\_to\_mpoly\_poly} in one of our main results regarding polynomials, which is useful in our correctness proof for \isa{elim\_forall} (cross reference \rref{sec:qe}):
\begin{isabelle}
\reindexeduniveval
\end{isabelle}

This lemma relates the evaluation of multivariate polynomials, of type \isa{real\ mpoly}, and multivariate polynomials \textit{treated as univariate polynomials} in the variable of interest \isa{Var 0}, of type \isa{rmpoly}.
To fully understand it, we must explain a few Isabelle/HOL operators that manipulate multivariate polynomials.
Here, \isa{eval\_mpoly} is our name for the natural definition of multivariate polynomial evaluation which substitutes real values for variables.
Because variables are represented with de Bruijn indices, we can store the values to substitute in a list \isa{L}, where the element of \isa{L} at position 0 is then substituted for \isa{Var\ 0}, the element of \isa{L} at position 1 is substituted for \isa{Var\ 1}, and so on.
If the length of \isa{L} is shorter than the number of variables, a default value of 0 is substituted for any variables that are not covered by \isa{L}.
This definition was implicitly used in prior work \cite{scharager2021verified}, but without being explicitly stated and named:
\begin{isabelle}
\evalmpoly
\end{isabelle}
The \isa{eval\_mpoly\_poly} function maps \isa{eval\_mpoly} over the coefficients of a \isa{real\ mpoly\ poly}.

Continuing to unpack the \isa{reindexed\_univ\_qs\_eval} lemma, the \isa{lowerPoly} function is from Scharager \emph{et al.} \cite{scharager2021verified}; here, it serves to reindex variables in multivariate polynomials, so that \isa{lowerPoly\ 0\ 1\ q} lowers every variable index in \isa{q} by 1.
The \isa{univariate\_in} operator is our function to perform this multivariate to univariate transformation.
Let $q_i$ be the polynomial at the $i$th index of \isa{qs}, and $uq_i$ be the polynomial at the $i$th index of \isa{univ\_qs}---then the first assumption of \isa{reindexed\_univ\_qs\_eval} says that $uq_i$ is the polynomial that we obtain by treating $q_i$ as univariate in \isa{Var 0}.

Next, the second assumption in \isa{reindexed\_univ\_qs\_eval} says that \isa{reindexed\_univ\_qs} is the list of polynomials obtained by lowering all variable indices in the \textit{coefficients} of the \isa{univ\_qs} by 1.
Let us call $ruq_i$ the polynomial at the $i$th index of  \isa{reindexed\_univ\_qs}.
Then, lemma \isa{reindexed\_univ\_qs\_eval} captures the mathematical equivalence of $q_i$ and $ruq_i$ by showing that evaluating $q_i$ on the valuation $v$ = \isa{x\isacharhash xs} gives the same result as evaluating the \textit{coefficients} of $ruq_i$ on \isa{xs} and then evaluating the resulting univariate polynomial (which now has constant coefficients) on \isa{x}.

The proof of this key lemma required that we first prove the following fundamental extensionality result, which says that if two polynomials \isa{p} and \isa{q} (in $n$ variables) have identical evaluations on $\mathbb{R}^n$, then they are themselves identical:
\begin{isabelle}
\mpolyeval
\end{isabelle}

Since real multivariate polynomials are fundamental to many areas of mathematics, it is our hope that our library developments will be useful to others, including in the formalization of other QE algorithms, but also more widely.

\subsection{Code Export}\label{sec:Export}
We export our multivariate QE algorithm to SML code, which removes overhead and allows us to better test our algorithm on examples.\footnote{This step requires trusting Isabelle/HOL's code generator in addition to the theorem prover's trusted core. Partial progress has been made on verifying Isabelle's code generator \cite{DBLP:conf/esop/HupelN18}.}
Building on the framework of Scharager \emph{et al.} (by using the same type for QE formulas and the same evaluation function for formulas) makes the connection with the verified virtual substitution algorithm \cite{scharager2021verified} very easy.\footnote{The top-level correctness theorems for verified virtual substitution \cite{scharager2021verified} have a very similar shape to \isa{qe\_correct}, as they state that for each top-level formalized virtual substitution method \isa{V} and valuation \isa{\isasymnu}, \isa{eval F \isasymnu} equals \isa{eval\ (V\ F)\ \isasymnu}.
This makes it easy to verify that, for any valuation \isa{\isasymnu}, \isa{eval\ F\ \isasymnu} equals \isa{eval\ ((qe \isasymcirc\ V)\ F)\ \isasymnu}.}
This means that we are able to retain efficiency \cite{scharager2021verified} on examples that are tractable for virtual substitution.

However, because virtual substitution is \textit{not} a complete QE method (i.e., it is not able to solve all QE problems), the efficiency, or lack thereof, of our (complete) algorithm is still significant.
Unfortunately (but not unexpectedly), without the link to virtual substitution, our hybrid multivariate algorithm is not at all efficient; it appears to hang on all but the simplest univariate examples.
However, we do not consider our algorithm's present inefficiency to be a fatal flaw, since we envision it as being a (major) stepping stone on the way towards an optimized algorithm.
As noted previously \cite{scharager2021verified}, unverified computer algebra systems have realized efficient QE in part because many have been extensively optimized over several decades; thus, it is natural that optimized verified algorithms will similarly take time to develop.

While inefficiency is not unexpected given that even Renegar may not realize practical efficiency in its current state \cite{HongTechRpt,DBLP:journals/cj/HeintzRS93}, at present, we suspect that part of the efficiency bottleneck for our algorithm is the untenable branching in the computation of the multivariate Tarski queries; this can be significantly reduced in the future by implementing an algorithm that more closely follows BKR.
We also believe that our algorithm's lack of inherent optimizations is another contributing factor; as one example, we currently branch unnecessarily on the signs of constant coefficients.
Further, we are not currently exploiting the algorithm's inherent parallelism.
However, it does not make sense to focus on optimizing our algorithm at this stage (optimizations may be brittle).
Once the branching reflects the full reduction of BKR, then inefficiencies (such as the unnecessary branching on constant coefficients) should be identified and handled appropriately.

\section{Related Work}
From a theoretical standpoint, the most closely related work is one by Cyril Cohen, who formalized a sign-determination algorithm with reduction in Coq that, to our understanding, uses the same matrix equation as our algorithm, although the details of his formalization look quite different from ours.\footnote{This is in part because the setup is considerably different: while we extended a univariate QE procedure with reduction into multivariate, Cohen added reduction to an already multivariate sign-determination procedure.}
To our knowledge, he has not yet used this improved sign-determination algorithm for a QE algorithm, and this work is unpublished, but a writeup is available on his webpage \cite{CyrilReduction}.
Additionally, because the algorithm we verify is a hybrid between Tarski's QE algorithm and BKR, our work shares some theoretical overlap with Cohen and Mahboubi's formalization of Tarski's algorithm in Coq \cite{cohen_phd,AssiaQE}.

From a practical standpoint, we benefit from the well-developed Isabelle/HOL libraries.
This includes, of course, our previous verification of univariate BKR \cite{BKR} and our verification of virtual substitution \cite{scharager2021verified}, which have already been discussed at length.
Additionally, we build on the formalization of pseudo-remainder sequences (recently made available on the AFP \cite{Sturm_Tarski-AFP}) described by Li, Passmore, and Paulson \cite{li2019deciding}.
Although we formalize our own functions to generate pseudo-remainder sequences, which interface well with our assumptions-based framework (and which are specialized to the \isa{rmpoly} type), we derive insights from Li's code and mimic some of his structure in our functions, adapted appropriately to our purposes.
We also benefit from proving a connection between our functions and his.

\section{Conclusion and Future Work}
We develop and formalize Isabelle/HOL's first complete multivariate quantifier elimination (QE) algorithm for the first-order logic of real arithmetic.
Our algorithm mixes ideas from Tarski's original QE algorithm \cite{Tarski} and more efficient algorithms by BKR \cite{DBLP:journals/jcss/Ben-OrKR86} and Renegar \cite{DBLP:journals/jsc/Renegar92b}; the formalization requires rigorizing high-level mathematical insights \cite{DBLP:journals/jcss/Ben-OrKR86,DBLP:journals/jsc/Renegar92b}.
We realize a number of ideas suggested in our prior work by extending a formalization of univariate BKR \cite{BKR} to the multivariate case and by building on the framework of Scharager \emph{et al.} \cite{scharager2021verified} in order to link our work with an efficient verified virtual substitution QE algorithm.
While our algorithm (on its own) currently has prohibitive inefficiency, its nontrivial library extensions and theoretical interest (including its potential to be extended into variant algorithms that have promising parallel complexity \cite{1993Improved,DBLP:journals/aaecc/CuckerLMPR92,DBLP:journals/jsc/Renegar92b}) make it a meaningful contribution.

Future work includes first extending our algorithm to one that realizes the full reduction of BKR \cite{DBLP:journals/jcss/Ben-OrKR86}.
After this, it would be interesting to identify other areas of inefficiency and aggressively optimize.
In addition to fine-tuning the branching to avoid splitting on trivial cases (most notably, on constants), one very significant (and challenging) task will be to optimize the computation of the Tarski queries; this was previously noted in the univariate case also \cite{BKR}.
Overall, our contribution lays considerable groundwork for more optimized verified QE algorithms with inherent parallelism.

\section*{Acknowledgments}
We thank the anonymous CPP reviewers for their helpful feedback on the paper.

This material is based upon work supported by the National Science Foundation under Grant No. CNS-1739629, a National Science Foundation Graduate Research Fellowship under Grants Nos. DGE1252522 and DGE1745016, by the AFOSR under grant number FA9550-16-1-0288, by A*STAR, Singapore, and the Alexander von Humboldt Foundation. Any opinions, findings, and conclusions or recommendations expressed in this material are those of the author(s) and do not necessarily reflect the views of the National Science Foundation, AFOSR, or A*STAR.

\balance
\bibliographystyle{ACM-Reference-Format}
\bibliography{cpp}


\begin{thebibliography}{42}

%%% the Web address.

\ifx \showCODEN    \undefined \def \showCODEN     #1{\unskip}     \fi
\ifx \showDOI      \undefined \def \showDOI       #1{#1}\fi
\ifx \showISBNx    \undefined \def \showISBNx     #1{\unskip}     \fi
\ifx \showISBNxiii \undefined \def \showISBNxiii  #1{\unskip}     \fi
\ifx \showISSN     \undefined \def \showISSN      #1{\unskip}     \fi
\ifx \showLCCN     \undefined \def \showLCCN      #1{\unskip}     \fi
\ifx \shownote     \undefined \def \shownote      #1{#1}          \fi
\ifx \showarticletitle \undefined \def \showarticletitle #1{#1}   \fi
\ifx \showURL      \undefined \def \showURL       {\relax}        \fi
% invisible to TeX
\providecommand\bibfield[2]{#2}
\providecommand\bibinfo[2]{#2}
\providecommand\natexlab[1]{#1}
\providecommand\showeprint[2][]{arXiv:#2}

\bibitem[Basu et~al\mbox{.}(2006)]%
        {algRAG}
\bibfield{author}{\bibinfo{person}{Saugata Basu}, \bibinfo{person}{Richard
  Pollack}, {and} \bibinfo{person}{Marie-Fran\c{c}oise Roy}.}
  \bibinfo{year}{2006}\natexlab{}.
\newblock \bibinfo{booktitle}{\emph{Algorithms in Real Algebraic Geometry}
  (\bibinfo{edition}{second} ed.)}.
\newblock \bibinfo{publisher}{Springer}, \bibinfo{address}{Berlin, Heidelberg}.
\newblock
\urldef\tempurl%
\url{https://doi.org/10.1007/3-540-33099-2}
\showDOI{\tempurl}


\bibitem[Ben{-}Or et~al\mbox{.}(1986)]%
        {DBLP:journals/jcss/Ben-OrKR86}
\bibfield{author}{\bibinfo{person}{Michael Ben{-}Or}, \bibinfo{person}{Dexter
  Kozen}, {and} \bibinfo{person}{John~H. Reif}.}
  \bibinfo{year}{1986}\natexlab{}.
\newblock \showarticletitle{The Complexity of Elementary Algebra and Geometry}.
\newblock \bibinfo{journal}{\emph{J. Comput. Syst. Sci.}} \bibinfo{volume}{32},
  \bibinfo{number}{2} (\bibinfo{year}{1986}), \bibinfo{pages}{251--264}.
\newblock
\urldef\tempurl%
\url{https://doi.org/10.1016/0022-0000(86)90029-2}
\showDOI{\tempurl}


\bibitem[Brown(2001)]%
        {Brown}
\bibfield{author}{\bibinfo{person}{Christopher~W. Brown}.}
  \bibinfo{year}{2001}\natexlab{}.
\newblock \showarticletitle{Improved Projection for Cylindrical Algebraic
  Decomposition}.
\newblock \bibinfo{journal}{\emph{J. Symb. Comput.}} \bibinfo{volume}{32},
  \bibinfo{number}{5} (\bibinfo{year}{2001}), \bibinfo{pages}{447--465}.
\newblock
\urldef\tempurl%
\url{https://doi.org/10.1006/jsco.2001.0463}
\showDOI{\tempurl}


\bibitem[Brown(2003)]%
        {DBLP:journals/cca/Brown03}
\bibfield{author}{\bibinfo{person}{Christopher~W. Brown}.}
  \bibinfo{year}{2003}\natexlab{}.
\newblock \showarticletitle{{QEPCAD} {B:} a program for computing with
  semi-algebraic sets using CADs}.
\newblock \bibinfo{journal}{\emph{{SIGSAM} Bull.}} \bibinfo{volume}{37},
  \bibinfo{number}{4} (\bibinfo{year}{2003}), \bibinfo{pages}{97--108}.
\newblock
\urldef\tempurl%
\url{https://doi.org/10.1145/968708.968710}
\showDOI{\tempurl}


\bibitem[Canny(1993)]%
        {1993Improved}
\bibfield{author}{\bibinfo{person}{John~F. Canny}.}
  \bibinfo{year}{1993}\natexlab{}.
\newblock \showarticletitle{Improved Algorithms for Sign Determination and
  Existential Quantifier Elimination}.
\newblock \bibinfo{journal}{\emph{Comput. J.}} \bibinfo{volume}{36},
  \bibinfo{number}{5} (\bibinfo{year}{1993}), \bibinfo{pages}{409--418}.
\newblock
\urldef\tempurl%
\url{https://doi.org/10.1093/comjnl/36.5.409}
\showDOI{\tempurl}


\bibitem[Cohen(2012)]%
        {cohen_phd}
\bibfield{author}{\bibinfo{person}{Cyril Cohen}.}
  \bibinfo{year}{2012}\natexlab{}.
\newblock \emph{\bibinfo{title}{{Formalized algebraic numbers: construction and
  first-order theory}}}.
\newblock \bibinfo{thesistype}{Ph.\,D. Dissertation}.
  \bibinfo{school}{{\'E}cole polytechnique}.
\newblock
\urldef\tempurl%
\url{https://perso.crans.org/cohen/papers/thesis.pdf}
\showURL{%
\tempurl}


\bibitem[Cohen(2021)]%
        {CyrilReduction}
\bibfield{author}{\bibinfo{person}{Cyril Cohen}.}
  \bibinfo{year}{2021}\natexlab{}.
\newblock \bibinfo{title}{Formalization of a sign determination algorithm in
  real algebraic geometry}.  (\bibinfo{year}{2021}).
\newblock
\newblock
\shownote{Preprint on webpage at https://hal.inria.fr/hal-03274013/document}.


\bibitem[Cohen and Mahboubi(2012)]%
        {AssiaQE}
\bibfield{author}{\bibinfo{person}{Cyril Cohen} {and} \bibinfo{person}{Assia
  Mahboubi}.} \bibinfo{year}{2012}\natexlab{}.
\newblock \showarticletitle{Formal proofs in real algebraic geometry: from
  ordered fields to quantifier elimination}.
\newblock \bibinfo{journal}{\emph{Log. Methods Comput. Sci.}}
  \bibinfo{volume}{8}, \bibinfo{number}{1} (\bibinfo{year}{2012}).
\newblock
\urldef\tempurl%
\url{https://doi.org/10.2168/LMCS-8(1:2)2012}
\showDOI{\tempurl}


\bibitem[Collins(1975)]%
        {Collins}
\bibfield{author}{\bibinfo{person}{George~E. Collins}.}
  \bibinfo{year}{1975}\natexlab{}.
\newblock \showarticletitle{Quantifier elimination for real closed fields by
  cylindrical algebraic decomposition}. In \bibinfo{booktitle}{\emph{Automata
  Theory and Formal Languages}} \emph{(\bibinfo{series}{LNCS},
  Vol.~\bibinfo{volume}{33})}, \bibfield{editor}{\bibinfo{person}{H.~Barkhage}}
  (Ed.). \bibinfo{publisher}{Springer}, \bibinfo{pages}{134--183}.
\newblock
\urldef\tempurl%
\url{https://doi.org/10.1007/3-540-07407-4_17}
\showDOI{\tempurl}


\bibitem[Collins and Hong(1991)]%
        {DBLP:journals/jsc/CollinsH91}
\bibfield{author}{\bibinfo{person}{George~E. Collins} {and} \bibinfo{person}{H.
  Hong}.} \bibinfo{year}{1991}\natexlab{}.
\newblock \showarticletitle{Partial Cylindrical Algebraic Decomposition for
  Quantifier Elimination}.
\newblock \bibinfo{journal}{\emph{J. Symb. Comput.}} \bibinfo{volume}{12},
  \bibinfo{number}{3} (\bibinfo{year}{1991}), \bibinfo{pages}{299--328}.
\newblock
\urldef\tempurl%
\url{https://doi.org/10.1016/S0747-7171(08)80152-6}
\showDOI{\tempurl}


\bibitem[Cordwell et~al\mbox{.}(2021a)]%
        {BKR_AFP}
\bibfield{author}{\bibinfo{person}{Katherine Cordwell},
  \bibinfo{person}{Yong~Kiam Tan}, {and} \bibinfo{person}{André Platzer}.}
  \bibinfo{year}{2021}\natexlab{a}.
\newblock \showarticletitle{The {BKR} Decision Procedure for Univariate Real
  Arithmetic}.
\newblock \bibinfo{journal}{\emph{Archive of Formal Proofs}}
  (\bibinfo{date}{April} \bibinfo{year}{2021}).
\newblock
\newblock
\shownote{\url{https://www.isa-afp.org/entries/BenOr_Kozen_Reif.html}, Formal
  proof development}.


\bibitem[Cordwell et~al\mbox{.}(2021b)]%
        {BKR}
\bibfield{author}{\bibinfo{person}{Katherine Cordwell},
  \bibinfo{person}{Yong~Kiam Tan}, {and} \bibinfo{person}{Andr{\'{e}}
  Platzer}.} \bibinfo{year}{2021}\natexlab{b}.
\newblock \showarticletitle{A Verified Decision Procedure for Univariate Real
  Arithmetic with the {BKR} Algorithm}. In \bibinfo{booktitle}{\emph{ITP}}
  \emph{(\bibinfo{series}{LIPIcs}, Vol.~\bibinfo{volume}{193})},
  \bibfield{editor}{\bibinfo{person}{Liron Cohen} {and} \bibinfo{person}{Cezary
  Kaliszyk}} (Eds.). \bibinfo{publisher}{Schloss Dagstuhl - Leibniz-Zentrum
  f{\"{u}}r Informatik}, \bibinfo{pages}{14:1--14:20}.
\newblock
\urldef\tempurl%
\url{https://doi.org/10.4230/LIPIcs.ITP.2021.14}
\showDOI{\tempurl}


\bibitem[Cucker et~al\mbox{.}(1992)]%
        {DBLP:journals/aaecc/CuckerLMPR92}
\bibfield{author}{\bibinfo{person}{Felipe Cucker}, \bibinfo{person}{Herv{\'{e}}
  Lanneau}, \bibinfo{person}{Bud Mishra}, \bibinfo{person}{Paul Pedersen},
  {and} \bibinfo{person}{Marie{-}Fran{\c{c}}oise Roy}.}
  \bibinfo{year}{1992}\natexlab{}.
\newblock \showarticletitle{{NC} Algorithms for Real Algebraic Numbers}.
\newblock \bibinfo{journal}{\emph{Appl. Algebra Eng. Commun. Comput.}}
  \bibinfo{volume}{3} (\bibinfo{year}{1992}), \bibinfo{pages}{79--98}.
\newblock
\urldef\tempurl%
\url{https://doi.org/10.1007/BF01387193}
\showDOI{\tempurl}


\bibitem[Davenport and Heintz(1988)]%
        {DBLP:journals/jsc/DavenportH88}
\bibfield{author}{\bibinfo{person}{James~H. Davenport} {and}
  \bibinfo{person}{Joos Heintz}.} \bibinfo{year}{1988}\natexlab{}.
\newblock \showarticletitle{Real Quantifier Elimination is Doubly Exponential.}
\newblock \bibinfo{journal}{\emph{J. Symb. Comput.}} \bibinfo{volume}{5},
  \bibinfo{number}{1/2} (\bibinfo{year}{1988}), \bibinfo{pages}{29--35}.
\newblock
\urldef\tempurl%
\url{https://doi.org/10.1016/S0747-7171(88)80004-X}
\showDOI{\tempurl}


\bibitem[de~Moura and Passmore(2013)]%
        {DBLP:conf/cade/MouraP13}
\bibfield{author}{\bibinfo{person}{Leonardo~Mendon{\c{c}}a de Moura} {and}
  \bibinfo{person}{Grant~Olney Passmore}.} \bibinfo{year}{2013}\natexlab{}.
\newblock \showarticletitle{Computation in Real Closed Infinitesimal and
  Transcendental Extensions of the Rationals}. In
  \bibinfo{booktitle}{\emph{CADE}} \emph{(\bibinfo{series}{LNCS},
  Vol.~\bibinfo{volume}{7898})}, \bibfield{editor}{\bibinfo{person}{Maria~Paola
  Bonacina}} (Ed.). \bibinfo{publisher}{Springer}, \bibinfo{pages}{178--192}.
\newblock
\urldef\tempurl%
\url{https://doi.org/10.1007/978-3-642-38574-2\_12}
\showDOI{\tempurl}


\bibitem[Dolzmann et~al\mbox{.}(2004)]%
        {DBLP:conf/issac/DolzmannSS04}
\bibfield{author}{\bibinfo{person}{Andreas Dolzmann}, \bibinfo{person}{Andreas
  Seidl}, {and} \bibinfo{person}{Thomas Sturm}.}
  \bibinfo{year}{2004}\natexlab{}.
\newblock \showarticletitle{Efficient projection orders for {CAD}}. In
  \bibinfo{booktitle}{\emph{ISSAC}}, \bibfield{editor}{\bibinfo{person}{Jaime
  Gutierrez}} (Ed.). \bibinfo{publisher}{{ACM}}, \bibinfo{pages}{111--118}.
\newblock
\urldef\tempurl%
\url{https://doi.org/10.1145/1005285.1005303}
\showDOI{\tempurl}


\bibitem[Eberl(2015)]%
        {DBLP:conf/cpp/Eberl15}
\bibfield{author}{\bibinfo{person}{Manuel Eberl}.}
  \bibinfo{year}{2015}\natexlab{}.
\newblock \showarticletitle{A Decision Procedure for Univariate Real
  Polynomials in {I}sabelle/{HOL}}. In \bibinfo{booktitle}{\emph{CPP}},
  \bibfield{editor}{\bibinfo{person}{Xavier Leroy} {and} \bibinfo{person}{Alwen
  Tiu}} (Eds.). \bibinfo{publisher}{{ACM}}, \bibinfo{pages}{75--83}.
\newblock
\urldef\tempurl%
\url{https://doi.org/10.1145/2676724.2693166}
\showDOI{\tempurl}


\bibitem[Eberl and Thiemann(2021)]%
        {Factor_Algebraic_Polynomial-AFP}
\bibfield{author}{\bibinfo{person}{Manuel Eberl} {and} \bibinfo{person}{René
  Thiemann}.} \bibinfo{year}{2021}\natexlab{}.
\newblock \showarticletitle{Factorization of Polynomials with Algebraic
  Coefficients}.
\newblock \bibinfo{journal}{\emph{Archive of Formal Proofs}}
  (\bibinfo{date}{November} \bibinfo{year}{2021}).
\newblock
\showISSN{2150-914x}
\newblock
\shownote{\url{https://isa-afp.org/entries/Factor_Algebraic_Polynomial.html},
  Formal proof development}.


\bibitem[Harrison(2007)]%
        {DBLP:conf/tphol/Harrison07}
\bibfield{author}{\bibinfo{person}{John Harrison}.}
  \bibinfo{year}{2007}\natexlab{}.
\newblock \showarticletitle{Verifying Nonlinear Real Formulas Via Sums of
  Squares}. In \bibinfo{booktitle}{\emph{TPHOLs}}
  \emph{(\bibinfo{series}{LNCS}, Vol.~\bibinfo{volume}{4732})},
  \bibfield{editor}{\bibinfo{person}{Klaus Schneider} {and}
  \bibinfo{person}{Jens Brandt}} (Eds.). \bibinfo{publisher}{Springer},
  \bibinfo{pages}{102--118}.
\newblock
\urldef\tempurl%
\url{https://doi.org/10.1007/978-3-540-74591-4_9}
\showDOI{\tempurl}


\bibitem[Heintz et~al\mbox{.}(1993)]%
        {DBLP:journals/cj/HeintzRS93}
\bibfield{author}{\bibinfo{person}{Joos Heintz},
  \bibinfo{person}{Marie{-}Fran{\c{c}}oise Roy}, {and} \bibinfo{person}{Pablo
  Solern{\'{o}}}.} \bibinfo{year}{1993}\natexlab{}.
\newblock \showarticletitle{On the Theoretical and Practical Complexity of the
  Existential Theory of Reals}.
\newblock \bibinfo{journal}{\emph{Comput. J.}} \bibinfo{volume}{36},
  \bibinfo{number}{5} (\bibinfo{year}{1993}), \bibinfo{pages}{427--431}.
\newblock
\urldef\tempurl%
\url{https://doi.org/10.1093/comjnl/36.5.427}
\showDOI{\tempurl}


\bibitem[Hong(1991)]%
        {HongTechRpt}
\bibfield{author}{\bibinfo{person}{Hoon Hong}.}
  \bibinfo{year}{1991}\natexlab{}.
\newblock \bibinfo{booktitle}{\emph{Comparison of Several Decision Algorithms
  for the Existential Theory of the Reals}}.
\newblock \bibinfo{type}{Technical Report}. \bibinfo{institution}{RISC}.
\newblock
\urldef\tempurl%
\url{https://citeseerx.ist.psu.edu/viewdoc/summary?doi=10.1.1.30.8707}
\showURL{%
\tempurl}


\bibitem[Hupel and Nipkow(2018)]%
        {DBLP:conf/esop/HupelN18}
\bibfield{author}{\bibinfo{person}{Lars Hupel} {and} \bibinfo{person}{Tobias
  Nipkow}.} \bibinfo{year}{2018}\natexlab{}.
\newblock \showarticletitle{A Verified Compiler from {Isabelle}/{HOL} to
  {CakeML}}. In \bibinfo{booktitle}{\emph{ESOP}} \emph{(\bibinfo{series}{LNCS},
  Vol.~\bibinfo{volume}{10801})}, \bibfield{editor}{\bibinfo{person}{Amal
  Ahmed}} (Ed.). \bibinfo{publisher}{Springer}, \bibinfo{pages}{999--1026}.
\newblock
\urldef\tempurl%
\url{https://doi.org/10.1007/978-3-319-89884-1\_35}
\showDOI{\tempurl}


\bibitem[Kosaian et~al\mbox{.}(2022)]%
        {Multiv_BKR_AFP}
\bibfield{author}{\bibinfo{person}{Katherine Kosaian},
  \bibinfo{person}{Yong~Kiam Tan}, {and} \bibinfo{person}{André Platzer}.}
  \bibinfo{year}{2022}\natexlab{}.
\newblock \showarticletitle{A First Complete Algorithm for Real Quantifier
  Elimination in {I}sabelle/{HOL}}.
\newblock \bibinfo{journal}{\emph{Archive of Formal Proofs}}
  (\bibinfo{date}{Dec.} \bibinfo{year}{2022}).
\newblock
\newblock
\shownote{\url{https://www.isa-afp.org/entries/Quantifier_Elimination_Hybrid.html},
  Formal proof development}.


\bibitem[Li(2014)]%
        {Sturm_Tarski-AFP}
\bibfield{author}{\bibinfo{person}{Wenda Li}.} \bibinfo{year}{2014}\natexlab{}.
\newblock \showarticletitle{The {S}turm-{T}arski Theorem}.
\newblock \bibinfo{journal}{\emph{Archive of Formal Proofs}}
  (\bibinfo{date}{Sept.} \bibinfo{year}{2014}).
\newblock
\showISSN{2150-914x}
\newblock
\shownote{\url{https://isa-afp.org/entries/Sturm_Tarski.html}, Formal proof
  development}.


\bibitem[Li et~al\mbox{.}(2019)]%
        {li2019deciding}
\bibfield{author}{\bibinfo{person}{Wenda Li}, \bibinfo{person}{Grant~Olney
  Passmore}, {and} \bibinfo{person}{Lawrence~C. Paulson}.}
  \bibinfo{year}{2019}\natexlab{}.
\newblock \showarticletitle{Deciding Univariate Polynomial Problems Using
  Untrusted Certificates in {Isabelle/HOL}}.
\newblock \bibinfo{journal}{\emph{J. Autom. Reason.}} \bibinfo{volume}{62},
  \bibinfo{number}{1} (\bibinfo{year}{2019}), \bibinfo{pages}{69--91}.
\newblock
\urldef\tempurl%
\url{https://doi.org/10.1007/s10817-017-9424-6}
\showDOI{\tempurl}


\bibitem[Li and Paulson(2016)]%
        {li2016modular}
\bibfield{author}{\bibinfo{person}{Wenda Li} {and} \bibinfo{person}{Lawrence~C.
  Paulson}.} \bibinfo{year}{2016}\natexlab{}.
\newblock \showarticletitle{A modular, efficient formalisation of real
  algebraic numbers}. In \bibinfo{booktitle}{\emph{CPP}},
  \bibfield{editor}{\bibinfo{person}{Jeremy Avigad} {and} \bibinfo{person}{Adam
  Chlipala}} (Eds.). \bibinfo{publisher}{{ACM}}, \bibinfo{pages}{66--75}.
\newblock
\urldef\tempurl%
\url{https://doi.org/10.1145/2854065.2854074}
\showDOI{\tempurl}


\bibitem[Mahboubi(2007)]%
        {DBLP:journals/mscs/Mahboubi07}
\bibfield{author}{\bibinfo{person}{Assia Mahboubi}.}
  \bibinfo{year}{2007}\natexlab{}.
\newblock \showarticletitle{Implementing the cylindrical algebraic
  decomposition within the Coq system}.
\newblock \bibinfo{journal}{\emph{Math. Struct. Comput. Sci.}}
  \bibinfo{volume}{17}, \bibinfo{number}{1} (\bibinfo{year}{2007}),
  \bibinfo{pages}{99--127}.
\newblock
\urldef\tempurl%
\url{https://doi.org/10.1017/S096012950600586X}
\showDOI{\tempurl}


\bibitem[McCallum(1985)]%
        {McCallumProj}
\bibfield{author}{\bibinfo{person}{Scott McCallum}.}
  \bibinfo{year}{1985}\natexlab{}.
\newblock \showarticletitle{An Improved Projection Operation for Cylindrical
  Algebraic Decomposition}. In \bibinfo{booktitle}{\emph{EUROCAL}}
  \emph{(\bibinfo{series}{LNCS}, Vol.~\bibinfo{volume}{204})},
  \bibfield{editor}{\bibinfo{person}{B.~F. Caviness}} (Ed.).
  \bibinfo{publisher}{Springer}, \bibinfo{pages}{277--278}.
\newblock
\urldef\tempurl%
\url{https://doi.org/10.1007/3-540-15984-3\_277}
\showDOI{\tempurl}


\bibitem[McLaughlin and Harrison(2005)]%
        {harrison}
\bibfield{author}{\bibinfo{person}{Sean McLaughlin} {and} \bibinfo{person}{John
  Harrison}.} \bibinfo{year}{2005}\natexlab{}.
\newblock \showarticletitle{A Proof-Producing Decision Procedure for Real
  Arithmetic}. In \bibinfo{booktitle}{\emph{CADE}}
  \emph{(\bibinfo{series}{LNCS}, Vol.~\bibinfo{volume}{3632})},
  \bibfield{editor}{\bibinfo{person}{Robert Nieuwenhuis}} (Ed.).
  \bibinfo{publisher}{Springer}, \bibinfo{pages}{295--314}.
\newblock
\urldef\tempurl%
\url{https://doi.org/10.1007/11532231_22}
\showDOI{\tempurl}


\bibitem[Mu{\~{n}}oz et~al\mbox{.}(2018)]%
        {NASAHutch}
\bibfield{author}{\bibinfo{person}{C{\'{e}}sar~A. Mu{\~{n}}oz},
  \bibinfo{person}{Anthony~J. Narkawicz}, {and} \bibinfo{person}{Aaron Dutle}.}
  \bibinfo{year}{2018}\natexlab{}.
\newblock \showarticletitle{A Decision Procedure for Univariate Polynomial
  Systems Based on Root Counting and Interval Subdivision}.
\newblock \bibinfo{journal}{\emph{J. Formaliz. Reason.}} \bibinfo{volume}{11},
  \bibinfo{number}{1} (\bibinfo{year}{2018}), \bibinfo{pages}{19--41}.
\newblock
\urldef\tempurl%
\url{https://doi.org/10.6092/issn.1972-5787/8212}
\showDOI{\tempurl}


\bibitem[Narkawicz et~al\mbox{.}(2015)]%
        {NASATarski}
\bibfield{author}{\bibinfo{person}{Anthony Narkawicz},
  \bibinfo{person}{C{\'{e}}sar~A. Mu{\~{n}}oz}, {and} \bibinfo{person}{Aaron
  Dutle}.} \bibinfo{year}{2015}\natexlab{}.
\newblock \showarticletitle{Formally-Verified Decision Procedures for
  Univariate Polynomial Computation Based on {Sturm}'s and {Tarski}'s
  Theorems}.
\newblock \bibinfo{journal}{\emph{J. Autom. Reason.}} \bibinfo{volume}{54},
  \bibinfo{number}{4} (\bibinfo{year}{2015}), \bibinfo{pages}{285--326}.
\newblock
\urldef\tempurl%
\url{https://doi.org/10.1007/s10817-015-9320-x}
\showDOI{\tempurl}


\bibitem[Nipkow(2010)]%
        {DBLP:journals/jar/Nipkow10}
\bibfield{author}{\bibinfo{person}{Tobias Nipkow}.}
  \bibinfo{year}{2010}\natexlab{}.
\newblock \showarticletitle{Linear Quantifier Elimination}.
\newblock \bibinfo{journal}{\emph{J. Autom. Reason.}} \bibinfo{volume}{45},
  \bibinfo{number}{2} (\bibinfo{year}{2010}), \bibinfo{pages}{189--212}.
\newblock
\urldef\tempurl%
\url{https://doi.org/10.1007/s10817-010-9183-0}
\showDOI{\tempurl}


\bibitem[Paulson and Blanchette(2010)]%
        {DBLP:conf/lpar/PaulsonB10}
\bibfield{author}{\bibinfo{person}{Lawrence~C. Paulson} {and}
  \bibinfo{person}{Jasmin~Christian Blanchette}.}
  \bibinfo{year}{2010}\natexlab{}.
\newblock \showarticletitle{Three years of experience with {S}ledgehammer, a
  Practical Link Between Automatic and Interactive Theorem Provers}. In
  \bibinfo{booktitle}{\emph{IWIL}} \emph{(\bibinfo{series}{EPiC Series in
  Computing}, Vol.~\bibinfo{volume}{2})},
  \bibfield{editor}{\bibinfo{person}{Geoff Sutcliffe}, \bibinfo{person}{Stephan
  Schulz}, {and} \bibinfo{person}{Eugenia Ternovska}} (Eds.).
  \bibinfo{publisher}{EasyChair}, \bibinfo{pages}{1--11}.
\newblock


\bibitem[Platzer(2018)]%
        {Platzer18}
\bibfield{author}{\bibinfo{person}{Andr{\'e} Platzer}.}
  \bibinfo{year}{2018}\natexlab{}.
\newblock \bibinfo{booktitle}{\emph{Logical Foundations of Cyber-Physical
  Systems}}.
\newblock \bibinfo{publisher}{Springer}, \bibinfo{address}{Cham}.
\newblock
\showISBNx{978-3-319-63587-3}
\urldef\tempurl%
\url{https://doi.org/10.1007/978-3-319-63588-0}
\showDOI{\tempurl}


\bibitem[Platzer et~al\mbox{.}(2009)]%
        {DBLP:conf/cade/PlatzerQR09}
\bibfield{author}{\bibinfo{person}{Andr{\'{e}} Platzer},
  \bibinfo{person}{Jan{-}David Quesel}, {and} \bibinfo{person}{Philipp
  R{\"{u}}mmer}.} \bibinfo{year}{2009}\natexlab{}.
\newblock \showarticletitle{Real World Verification}. In
  \bibinfo{booktitle}{\emph{CADE}} \emph{(\bibinfo{series}{LNCS},
  Vol.~\bibinfo{volume}{5663})}, \bibfield{editor}{\bibinfo{person}{Renate~A.
  Schmidt}} (Ed.). \bibinfo{publisher}{Springer}, \bibinfo{pages}{485--501}.
\newblock
\urldef\tempurl%
\url{https://doi.org/10.1007/978-3-642-02959-2_35}
\showDOI{\tempurl}


\bibitem[Renegar(1992)]%
        {DBLP:journals/jsc/Renegar92b}
\bibfield{author}{\bibinfo{person}{James Renegar}.}
  \bibinfo{year}{1992}\natexlab{}.
\newblock \showarticletitle{On the Computational Complexity and Geometry of the
  First-Order Theory of the Reals, Part {III:} {Q}uantifier Elimination}.
\newblock \bibinfo{journal}{\emph{J. Symb. Comput.}} \bibinfo{volume}{13},
  \bibinfo{number}{3} (\bibinfo{year}{1992}), \bibinfo{pages}{329--352}.
\newblock
\urldef\tempurl%
\url{https://doi.org/10.1016/S0747-7171(10)80005-7}
\showDOI{\tempurl}


\bibitem[Scharager et~al\mbox{.}(2021a)]%
        {scharager2021verified}
\bibfield{author}{\bibinfo{person}{Matias Scharager},
  \bibinfo{person}{Katherine Cordwell}, \bibinfo{person}{Stefan Mitsch}, {and}
  \bibinfo{person}{Andr{\'{e}} Platzer}.} \bibinfo{year}{2021}\natexlab{a}.
\newblock \showarticletitle{Verified Quadratic Virtual Substitution for Real
  Arithmetic}. In \bibinfo{booktitle}{\emph{FM}} \emph{(\bibinfo{series}{LNCS},
  Vol.~\bibinfo{volume}{13047})}, \bibfield{editor}{\bibinfo{person}{Marieke
  Huisman}, \bibinfo{person}{Corina~S. Pasareanu}, {and}
  \bibinfo{person}{Naijun Zhan}} (Eds.). \bibinfo{publisher}{Springer},
  \bibinfo{pages}{200--217}.
\newblock
\urldef\tempurl%
\url{https://doi.org/10.1007/978-3-030-90870-6\_11}
\showDOI{\tempurl}


\bibitem[Scharager et~al\mbox{.}(2021b)]%
        {Virtual_Substitution-AFP}
\bibfield{author}{\bibinfo{person}{Matias Scharager},
  \bibinfo{person}{Katherine Cordwell}, \bibinfo{person}{Stefan Mitsch}, {and}
  \bibinfo{person}{André Platzer}.} \bibinfo{year}{2021}\natexlab{b}.
\newblock \showarticletitle{Verified Quadratic Virtual Substitution for Real
  Arithmetic}.
\newblock \bibinfo{journal}{\emph{Archive of Formal Proofs}}
  (\bibinfo{date}{October} \bibinfo{year}{2021}).
\newblock
\showISSN{2150-914x}
\newblock
\shownote{\url{https://isa-afp.org/entries/Virtual_Substitution.html}, Formal
  proof development}.


\bibitem[Strzebo{\'{n}}ski(2000)]%
        {strzeMathematica}
\bibfield{author}{\bibinfo{person}{Adam Strzebo{\'{n}}ski}.}
  \bibinfo{year}{2000}\natexlab{}.
\newblock \showarticletitle{Solving algebraic inequalities}.
\newblock \bibinfo{journal}{\emph{The Mathematica Journal}}
  \bibinfo{volume}{7}, \bibinfo{number}{4} (\bibinfo{year}{2000}),
  \bibinfo{pages}{525–541}.
\newblock


\bibitem[Tarski(1951)]%
        {Tarski}
\bibfield{author}{\bibinfo{person}{Alfred Tarski}.}
  \bibinfo{year}{1951}\natexlab{}.
\newblock \bibinfo{booktitle}{\emph{A Decision Method for Elementary Algebra
  and Geometry}}.
\newblock \bibinfo{publisher}{RAND Corporation}, \bibinfo{address}{Santa
  Monica, CA}.
\newblock
\urldef\tempurl%
\url{https://www.rand.org/pubs/reports/R109.html}
\showURL{%
\tempurl}


\bibitem[Weispfenning(1988)]%
        {DBLP:journals/jsc/Weispfenning88}
\bibfield{author}{\bibinfo{person}{Volker Weispfenning}.}
  \bibinfo{year}{1988}\natexlab{}.
\newblock \showarticletitle{The Complexity of Linear Problems in Fields}.
\newblock \bibinfo{journal}{\emph{J. Symb. Comput.}} \bibinfo{volume}{5},
  \bibinfo{number}{1-2} (\bibinfo{year}{1988}), \bibinfo{pages}{3--27}.
\newblock
\urldef\tempurl%
\url{https://doi.org/10.1016/S0747-7171(88)80003-8}
\showDOI{\tempurl}


\bibitem[Wenzel(2006)]%
        {DBLP:conf/mkm/Wenzel06}
\bibfield{author}{\bibinfo{person}{Makarius Wenzel}.}
  \bibinfo{year}{2006}\natexlab{}.
\newblock \showarticletitle{Structured Induction Proofs in {Isabelle/Isar}}. In
  \bibinfo{booktitle}{\emph{MKM}} \emph{(\bibinfo{series}{LNCS},
  Vol.~\bibinfo{volume}{4108})}, \bibfield{editor}{\bibinfo{person}{Jonathan~M.
  Borwein} {and} \bibinfo{person}{William~M. Farmer}} (Eds.).
  \bibinfo{publisher}{Springer}, \bibinfo{pages}{17--30}.
\newblock
\urldef\tempurl%
\url{https://doi.org/10.1007/11812289\_3}
\showDOI{\tempurl}


\end{thebibliography}

\end{document}